\documentclass[prodmode,acmtomacs]{acmsmall} 

\usepackage[linesnumbered,ruled]{algorithm2e}

\SetAlFnt{\small}
\SetAlCapFnt{\small}
\SetAlCapNameFnt{\small}
\SetAlCapHSkip{0pt}
\IncMargin{-\parindent}
\NoCaptionOfAlgo

\acmVolume{0}
\acmNumber{0}
\acmArticle{00}
\acmYear{0000}
\acmMonth{1}
\usepackage{amssymb}
\usepackage{amsmath}
\usepackage{enumerate,array,multirow}
\usepackage{graphicx}
\usepackage{enumerate,caption,verbatim}
\usepackage{slashbox}
\usepackage{soul}
\newtheorem{assumption}{Assumption}

\begin{document}

\markboth{D. Ghoshdastidar, A. Dukkipati and S. Bhatnagar}{Smoothed Functional Algorithms using $q$-Gaussian Distributions}

\title{Smoothed Functional Algorithms for Stochastic Optimization using $q$-Gaussian Distributions}
\author{Debarghya Ghoshdastidar
\affil{Indian Institute of Science}
Ambedkar Dukkipati
\affil{Indian Institute of Science}
Shalabh Bhatnagar
\affil{Indian Institute of Science}}

\begin{abstract}
Smoothed functional (SF) schemes for gradient estimation are known to be
efficient in stochastic optimization algorithms, specially when the objective is to 
improve the performance of a stochastic system. However, the performance of 
these methods depends on several parameters, such as the choice of a suitable 
smoothing kernel. Different kernels have been studied in literature, which
include Gaussian, Cauchy and uniform distributions among others.
This paper studies a new class of kernels based on the $q$-Gaussian distribution, 
that has gained popularity in statistical physics over the last decade. 
Though the importance of this family of distributions is attributed to its ability 
to generalize the Gaussian distribution, we observe that this class encompasses
almost all existing smoothing kernels.
This motivates us to study SF schemes for gradient estimation
using the $q$-Gaussian distribution.
Using the derived gradient estimates, we propose two-timescale algorithms 
for optimization of a stochastic objective function in a constrained setting 
with projected gradient search approach. 
We prove the convergence of our algorithms to the set of stationary points of an associated ODE.
We also demonstrate their performance numerically through simulations on a queuing model. 
\end{abstract}




\acmformat{Debarghya Ghoshdastidar, Ambedkar Dukkipati and Shalabh Bhatnagar, 2013. 
Smoothed Functional Algorithms for Stochastic Optimization using $q$-Gaussian Distributions.}

\begin{bottomstuff}
Authors' address: 
D. Ghoshdastidar {and} A. Dukkipati {and} S. Bhatnagar,
Department of Computer Science \& Automation,
Indian Institute of Science, Bangalore - 560012, India.
\\Email: \{debarghya.g,ad,shalabh\}@csa.iisc.ernet.in
\end{bottomstuff}

\maketitle

\section{Introduction}

Optimization problems, where the objective function does not have an analytic expression,
are quite common in engineering and financial world. 
Such problems are often encountered in discrete event simulations
such as queuing systems, allocation problems or financial forecasting.
In many of these scenarios, the data, obtained via statistical survey or simulation,
contains only noisy estimates of the objective function to be optimized,
and hence, one is compelled to resort to stochastic techniques.
One of the most commonly used solution methodologies involves stochastic approximation algorithms, originally 
due to~\citeN{Robbins_1951_jour_AnnMathStat}, that is used to find the zeros of a given function.
Based on this approach, gradient descent algorithms have been developed,
in which the parameters controlling the system track the zeros of the gradient of the objective. However,
these algorithms require an estimate of the cost gradient.
One can employ direct gradient estimation techniques~\cite{Fu_2006_conf_Simulation} such as infinitesimal perturbation
analysis (IPA)~\cite{Suri_1987_jour_JACM}, which require problem-specific analysis,
and hence, have limited applicability.
In the general setting, \citeN{Kiefer_1952_jour_AnnMathStat} provide a gradient estimate that requires a number of parallel
simulations of the system linear in the dimension of the parameter.
More efficient techniques for gradient estimation,  have been developed
based on the smoothed functional (SF) approach~\cite{Katkovnik_1972_jour_Automation},
simultaneous perturbation stochastic approximation (SPSA)~\cite{Spall_1992_jour_AutoControlTrans},
likelihood ratio~\cite{LEcuyer_1994_jour_ManSc} etc.
Stochastic variations of Newton-based optimization methods,
also known as adaptive Newton-based schemes have also been studied in the
literature~\cite{Ruppert_1985_jour_AnnStat,Spall_2000_jour_AutoCtrl,Bhatnagar_2007_jour_TOMACS}.

When the above schemes for gradient estimation are employed in optimization
methods involving the long-run average cost objective, the time complexity of the algorithms increases as 
the long-run average cost needs to be estimated after each parameter update.
A more efficient approach is to simultaneously perform 
the long-run averaging and parameter updates using different step-size schedules.
This class of algorithms constitute the multi-timescale 
stochastic approximation algorithms~\cite{Bhatnagar_1998_jour_ProbEnggInfoSc}.
Two-timescale gradient based optimization algorithms have also been developed
using SPSA and SF schemes~\cite{Bhatnagar_2003_jour_Simulation}.

In \cite{Bhatnagar_2007_jour_TOMACS}, the performance of
two-timescale gradient SF schemes based on Gaussian perturbations is studied 
on a queuing system. The results presented there indicate that the performance of the SF algorithms can be 
significantly improved by tuning a number of parameters, such as
the variance of the Gaussian distribution and the step-sizes.
In addition to these parameters, it is also known that improved performance can be attained 
when the perturbations follow other distributions, such as Cauchy distribution~\cite{Styblinski_1990_jour_NeuNet}.
In fact, the general theory of SF methods indicate that a variety of distributions can be 
used to construct smoothed functionals as long as they satisfy certain conditions~\cite{Rubinstein_1981_book_Wiley}.
A number of smoothing kernels have been studied in the
literature~\cite{Rubinstein_1981_book_Wiley,Kreimer_1988_jour_SIAMNumeAnal,Kreimer_1992_jour_AnnOR,Styblinski_1990_jour_NeuNet}.

\subsection*{Summary of our contributions}

The goal of this paper is to propose a new class of smoothing kernels
that encompasses most of the aforementioned kernels as special cases. 
This class of kernels is based on the $q$-Gaussian distributions, that
has been studied extensively in the field of nonextensive statistical 
mechanics~\cite{Prato_1999_jour_PhyRevE,Vignat_2007_jour_PhyA}.

One of our main contributions is to 
show that the $q$-Gaussian family of multivariate distributions 
satisfy the sufficient conditions for smoothing kernels discussed in~\cite{Rubinstein_1981_book_Wiley}. 
This allows us to work with a larger class of distributions in SF algorithms, in a unified way, where
the `shape parameter'  of the $q$-Gaussian controls behavior of the distribution. 
This parameter also controls the smoothness of the convolution,
thereby providing additional tuning in the form of the appropriate smoothing kernel,
that can coincide with the currently known smoothing kernels (listed in Table~\ref{tab_kernel}).

We derive for the first time,
smoothed functional algorithms with $q$-Gaussian perturbations,
that exhibit a power-law nature in certain cases. 
We present estimators for gradient of a function using the $q$-Gaussian smoothing kernel.
We also propose multi-timescale algorithms for stochastic optimization using $q${-}Gaussian based SF
that incorporate gradient  based search procedures, 
and prove the convergence of the proposed algorithms to the neighborhood of a local optimum.
The convergence analysis presented in this paper differs from the approaches
that have been studied earlier~\cite{Bhatnagar_2007_jour_TOMACS}.
Here, we provide a more straightforward technique using standard results
from~\cite{Borkar_2008_book_Cambridge,Kushner_1978_book_Springer}.
Furthermore, we perform simulations on a queuing network to illustrate the benefits of 
the $q$-Gaussian based SF algorithms compared to their Gaussian counterparts.
A shorter version of this paper containing only the one-simulation 
$q$-Gaussian SF algorithm, and without the convergence proof,
has been presented in~\cite{Ghoshdastidar_2012_conf_ISIT}.

The rest of the paper is organized as follows. The framework
for the optimization problem and some preliminaries on SF and $q$-Gaussians are
presented in Section~\ref{preliminaries}. 
Section~\ref{qGrad} validates the use of $q$-Gaussian as a smoothing kernel,
and presents gradient based algorithms using $q${-}Gaussian SF.
The convergence analysis of the proposed algorithms is discussed in Section~\ref{GqSF_convergence}.
Section~\ref{sim_results} presents simulations based on a numerical setting.
Section~\ref{conclusion} provides the concluding remarks.
The appendix describes a sampling technique for multivariate $q$-Gaussians
that is used in the proposed algorithms.

\section{Background and Preliminaries} 
\label{preliminaries}

\subsection{Problem Framework}

Let {$\{Y_n{(\theta)}:{n\in\mathbb{N}}\} \subset\mathbb{R}^d$} be a parameterized Markov process, depending on a tunable
parameter {$\theta\in C$}, where {$C$} is a compact and convex subset of {$\mathbb{R}^N$}. Let {${P}_{\theta}(x,\:\mathrm{d}y)$} 
denote the transition kernel of {$\{Y_n({\theta})\}$}. 
We assume the following.
\begin{assumption}
\label{ergodic}
For any fixed parameter $\theta\in C$,
the process {$\{Y_n({\theta})\}$} is ergodic and has a unique invariant measure $\nu_{\theta}$. 
\end{assumption}
We consider a Lipschitz continuous cost function 
{$h:\mathbb{R}^d\mapsto\mathbb{R}^+\bigcup \{0\}$} associated with the process.
Our objective is to choose an appropriate $\theta \in C$ in order to
minimize the long-run average cost
\begin{equation}
\label{Jdefn}
J(\theta) = \lim_{L\to\infty}\frac{1}{L} \mathsf{E} \bigg[\sum_{m=0}^{L-1}h\left(Y_m({\theta})\right) \bigg]
= \int\limits_{\mathbb{R}^d}h(y)\nu_{\theta}(\mathrm{d}y) \:,
\end{equation}
where $\mathsf{E}[\cdot]$ denotes the expectation of `$~\cdot~$'. 
The existence of the above limit is assured by Assumption~\ref{ergodic}
and the fact that $h$ is continuous, hence measurable. 
We use a gradient search procedure to optimize the average cost $J(\theta)$, 
and hence, the following technical requirement is necessary.
\begin{assumption}
\label{differentiable}
The function {$J(.)$} is continuously differentiable for all {$\theta\in C$}.
\end{assumption}
The above assumption provides the existence of $\nabla_{\theta} J(\theta)$,
which we will estimate via the smoothed functionals.
However, verification of the above assumption depends on the underlying process
and is non-trivial in most cases. One can observe that under certain conditions
(for instance when cost function $h(\cdot)$ is bounded),
Assumption~\ref{differentiable} can be translated to impose the condition of continuous
differentiability of the stationary measure $\nu_\theta$ for all $\theta\in C$.
This, in turn, would depend on a similar condition on the transition kernel $P_\theta(x,\mathrm{d}y)$.
Discussions on such conditions for finite state Markov processes can be found 
in~\cite{Schweitzer_1968_jour_AppProb}, and similar 
results for general state systems were presented in~\cite{Vazquez_1992_jour_AppProb}.
However, in the general case, such conditions are difficult to verify.
In addition to above, we also assume the existence of a stochastic Lyapunov function. 
This requires the notion of a non-anticipative sequence, defined below.

\begin{definition}
A random sequence of parameter vectors, {$(\theta_n)_{n\geqslant0} \subset C$}, controlling a process
{$\{Y_n({\theta_n})\}\subset\mathbb{R}^d$}, is said to be non-anticipative if the conditional probability
{$P(Y_{n+1}({\theta_{n+1}})\in\:\mathrm{d}y|\mathcal{F}_n) = {P}_{\theta_n} (Y_n({\theta_n}), \:\mathrm{d}y)$} 
almost surely for {$n\geqslant0$} and all Borel sets {$\mathrm{d}y\subset\mathbb{R}^d$},
where {$\mathcal{F}_n = \sigma(\theta_m,Y_m({\theta_m}),m\leqslant n)$}, {$n\geqslant0$} are the associated {$\sigma$}-fields. 
\end{definition}
It can be seen that the sequence of parameters obtained using the algorithms proposed in this paper form a non-anticipative sequence.

\begin{assumption}
\label{lyapunov}
Let {$(\theta_n)$} be a non-anticipative sequence of random parameters controlling the process {$\{Y_n(\theta_n)\}$},
and {$\mathcal{F}_n = \sigma(\theta_m,$} {$Y_m(\theta_m),m\leqslant n)$}, {$n\geqslant0$} be a sequence of associated {$\sigma$}-fields. 
There exists {$\epsilon_0>0$}, a compact set {$\mathcal{K}\subset\mathbb{R}^d$}, and a continuous function 
{$V:\mathbb{R}^d\mapsto\mathbb{R}^+\bigcup\{0\}$}, with
{$\lim_{\Vert{x}\Vert\to\infty} V(x) = \infty$}, such that 
\begin{enumerate}[(i)]
\item 
{$\sup\limits_n \mathsf{E}[V(Y_n)^2] < \infty$}, and
\item 
{$\mathsf{E}[V(Y_{n+1})|\mathcal{F}_n] \leqslant V(Y_n) - \epsilon_0$}, whenever {$Y_n\notin\mathcal{K}$}, {$n\geqslant0$}.
\end{enumerate} 
\end{assumption}
One may note that the definition of the function $V$ does not depend on the parameter sequence.
Assumption~\ref{lyapunov} ensures that the process under a tunable parameter remains stable,
and is not required, for instance, if  the single-stage cost function $h$ is bounded.

\subsection{Smoothed Functionals}
\label{preliminaries_SF}

We present the idea behind the smoothed functional approach proposed by \citeN{Katkovnik_1972_jour_Automation}. 
For a function $f:\mathbb{R}^N\mapsto\mathbb{R}$,
define $S_{\beta}^1 f:C\mapsto\mathbb{R}$ as
\begin{equation}
\label{SF1}
(S_{\beta}^{1}f)(\theta) = \int_{\mathbb{R}^N}G_{\beta}(\eta)f(\theta-\eta)\:\mathrm{d}\eta
 = \int_{\mathbb{R}^N}G_{\beta}(\theta-\eta)f(\eta)\:\mathrm{d}\eta,
\end{equation}
for all $\theta\in C$, 
provided the integral exists.
Here {$G_{\beta}:\mathbb{R}^N\mapsto\mathbb{R}$}
is called a smoothing kernel. The map $S_\beta^1$ 
is called a smoothed functional (SF), where the parameter {$\beta\in\mathbb{R}$} 
controls the smoothing properties of $S_\beta^{1}$.
Technically, one may allow the domain of $S_\beta^1 f$ to be $\mathbb{R}^N$, 
but for the purpose of this paper, we restrict the domain to $C$.
The superscript 1 has been used to indicate that $S_\beta^1$ is a one-sided SF.
A two-sided form of SF also exists, defined as
\begin{align}
(S_{\beta}^2f)(\theta) 
&= \frac{1}{2}\int_{\mathbb{R}^N}G_{\beta}(\eta)\big(f(\theta-\eta)+f(\theta+\eta)\big)\:\mathrm{d}\eta \;.
\label{SF2}
\end{align}

The idea behind using smoothed functionals is that if {$f$} is not well-behaved,
then {$S_{\beta}^i f$} is ``better-behaved'' (where $i=1,2$), \textit{i.e.},
it is easier to compute the derivative of $S_{\beta}^if$. 
Hence, one wishes $S_\beta^i$ to be differentiable, and also
retain the desirable properties of $f$, such as convexity (if $f$ is also convex).
\citeN{Rubinstein_1981_book_Wiley} established that the SF achieves the aforementioned properties if the kernel 
function satisfies the following: 
\begin{enumerate}[(P1)]
\item $G_\beta:\mathbb{R}^N\mapsto\mathbb{R}$ such that 
{$G_{\beta}(\eta) = \frac{1}{{\beta}^N}G_1\left(\frac{\eta}{\beta}\right)$} for all $\eta\in\mathbb{R}^N$, 
\item {$G_{\beta}(\eta)$} is piecewise differentiable in {$\eta$},
\item {$G_{\beta}(\eta)$} is a probability distribution function, \textit{i.e.}, 
{$(S_{\beta}^i f)(\theta)$} can be written as an expectation for $i=1,2$,
\item {$\lim_{\beta\to0}G_{\beta}(\eta) = \delta(\eta)$}, where {$\delta(\eta)$} is the Dirac delta function, and
\item {$\lim_{\beta\to0} (S_{\beta}^if)(\theta) = f(\theta)$} for $i=1,2$.
\end{enumerate}
The quantity $G_1(\cdot)$ in (P1) is $G_\beta(\cdot)$ evaluated at $\beta=1$.
The Gaussian distribution satisfies the above conditions, and has been used as a smoothing kernel 
\cite{Katkovnik_1972_jour_Automation,Styblinski_1990_jour_NeuNet}.
Here, in the multivariate ($N$-dimensional) case, the smoothing kernel $G_\beta$ is given 
by a $N$-dimensional Gaussian distribution with zero mean and covariance matrix 
$\beta^2I_{N\times N}$. The control of $\beta$ over the smoothing effect can be intuitively
seen as for lower value of $\beta$, the distribution is concentrated about its mean,
and hence, $S_\beta^i f$ is close to $f$. As $\beta$ increases, 
$S_\beta^i f$ tends to become more smooth as local fluctuations are averaged.

The SF approach provides an useful method for estimating the gradient of any function $f$
as discussed in~\cite{Rubinstein_1981_book_Wiley}. For the particular case of
Gaussian smoothing, the gradient of $S_\beta^1 f$ can be derived 
by taking the derivative on $G_\beta(\theta-\eta)$.
This interchange of derivative and integral is ensured by Leibnitz rule and property (P2).
A simple change of variables shows that the gradient can be written as
\begin{align}
 \nabla_\theta (S_\beta^1 f)(\theta)
 = \frac{1}{\beta}\mathsf{E}_{G_1(\eta)} [\eta f(\theta+\beta\eta)].
\end{align}

\citeN{Bhatnagar_2003_jour_Simulation} used this approach to estimate the gradient 
of the long-run average cost $J$ in terms of the single-stage cost $h$.
The authors also showed that for small $\beta$, $\nabla_\theta S_\beta^1 J$ and $\nabla_\theta J$
are close. Thus, the gradient estimator is of the form
\begin{equation}
\nabla_{\theta}J(\theta)\approx\frac{1}{\beta ML}\sum_{n=0}^{M-1}\sum_{m=0}^{L-1}{\eta_nh(Y_m(\theta_n+\beta\eta_n))}
\label{G_estimate1}
\end{equation}
for large $M$, $L$ and small {$\beta$},
where for each $n$, the parameter of the process {$\{Y_m(\theta_n+\beta\eta_n)\}_{m\geqslant0}$} 
includes a standard Gaussian perturbation ($\eta_n$) of the tuning parameter {$\theta_n\in C$}.
A two-timescale argument (discussed in Section~\ref{qGrad}) allows
us to use a single simulation of the process for obtaining~\eqref{G_estimate1}.
A similar two-simulation gradient estimator, based on the two-sided SF~\eqref{SF2},
was derived by \citeN{Bhatnagar_2007_jour_TOMACS} as
\begin{equation}
\label{G_estimate2}
\nabla_{\theta}J(\theta)\approx
\frac{1}{2\beta ML}\sum_{n=0}^{M-1}\sum_{m=0}^{L-1}\eta_n\big(h(Y_m^1(\theta_n+\beta\eta_n))-h(Y_m^2(\theta_n-\beta\eta_n))\big)
\end{equation}
for large $M$, $L$ and small {$\beta$}, where the two processes 
{$Y_m^1(\theta_n+\beta\eta_n)$} and {$Y_m^2(\theta_n-\beta\eta_n)$} are simulated in parallel.

\begin{table}[b]
\centering
\setlength{\extrarowheight}{4pt}
\begin{tabular}{|c|m{0.3\textwidth}|c|c|}
\hline
\multicolumn{2}{|c|}{Smoothing kernel} & Distribution, $G_\beta(\eta)$ & From $q$-Gaussian \\
\hline \hline
\multirow{4}{*}{\rotatebox[origin=c]{90}{\small Multivariate~~~}}
& Gaussian \cite{Katkovnik_1972_jour_Automation}
& $\frac{1}{(2\pi)^{N/2}\beta^{N}}\exp\left(-\frac{\Vert\eta\Vert^2}{2\beta^2}\right)$
& $q\to1$
\\ \cline{2-4}
& Cauchy \cite{Styblinski_1990_jour_NeuNet}
& $\frac{\Gamma(\frac{N+1}{2})}{\pi^{(N+1)/2}\beta^{N}}\big(1+\frac{\Vert\eta\Vert^2}{\beta^2}\big)^{-\frac{N+1}{2}}$
& $q=\big(1+\frac{2}{N+1}\big)$
\\ \cline{2-4}
& Kernel in random search \cite{Rubinstein_1981_book_Wiley}
& $\frac{1}{\beta^{N}}\left(1-\frac{\Vert\eta\Vert^2}{\beta^2}\right)_+$
& $q=0$
\\ \hline
\multirow{2}{*}{\rotatebox[origin=c]{90}{\small Univariate~~~~}}
& Uniform \cite{Kiefer_1952_jour_AnnMathStat}
& 
$\frac{1}{2\beta}
\mathbf{1}\left\{\eta\in[-\beta,\beta]\right\}$
& $q\to-\infty$
\\ \cline{2-4}
&Dirac delta \cite{Kreimer_1988_jour_SIAMNumeAnal}
& $\delta\left({\eta}\right)$
& $q\to-\infty, \beta\to0$
\\ \cline{2-4}
&Beta \cite{Kreimer_1992_jour_AnnOR}
& $\frac{\Gamma(a+b)(\beta+\eta)_+^{a-1}(\beta-\eta)_+^{b-1}
}{\Gamma(a)\Gamma(b)(2\beta)^{a+b-1}}$
& $q=\big(1-\frac{1}{a-1}\big)$ if $a=b$
\\ \hline
\end{tabular}
\caption{List of smoothing kernels studied in literature. 
Here, $\mathbf{1}\{\cdot\}$ is the indicator function, $\Gamma(\cdot)$ denotes Gamma function, and $y_+ := \max(y,0)$.}
\label{tab_kernel}
\end{table}

However, the Gaussian distribution is not the only smoothing kernel.
Few other alternatives that have been studied in the literature are listed in Table~\ref{tab_kernel}.
The last column of the table indicates
the condition under which these kernels can be retrieved as special cases of $q$-Gaussian distribution,
introduced in the next section.
The table provides a clear motivation for studying the smoothing properties 
of $q$-Gaussians as we can obtain most of the existing kernels through a suitable choice of $q$.
In addition, we can have infinitely many kernels at our disposal by 
tuning $q$, which is a continuous real-valued parameter.
To the the best of our knowledge, this is the first work on comparison of
existing smoothing kernels.

\subsection{The multivariate $q$-Gaussian distribution}

The $q$-Gaussian class of distributions was developed to describe the process 
of L\'{e}vy super-diffusion~\cite{Prato_1999_jour_PhyRevE}, but has been later studied in other 
fields, such as finance~\cite{Sato_2001_jour_JourPhyConf} and statistics~\cite{Suyari_2005_jour_ITTrans}.
The origin of this distribution is associated with the nonextensive generalization 
of Shannon entropy, introduced by~\citeN{Tsallis_1988_jour_StatPhy} in thermodynamics.
It results from maximizing Tsallis entropy 
under certain `deformed' moment constraints, known as normalized $q$-expectation~\cite{Tsallis_1998_jour_PhysicaA}.
This form of an expectation has been shown
to be compatible with the foundations of nonextensive statistics, and
it  coincides with the usual notion of expectation in the limiting
case of $q\to1$, a situation when Shannon entropy is also retrieved from Tsallis entropy.

Following the fact that the Gaussian distribution maximizes Shannon entropy
under mean and variance constraints,
\citeN{Prato_1999_jour_PhyRevE} maximized Tsallis entropy under the constraints,
\begin{equation}
\label{q-expect-defn}
\frac{\int_{\mathbb{R}}x p(x)^q\:\mathrm{d}x}{\int_{\mathbb{R}}p(x)^q\:\mathrm{d}x} = \mu_q
\qquad\text{ and } \qquad
\frac{\int_{\mathbb{R}}(x-\mu_q)^2 p(x)^q\:\mathrm{d}x}{\int_{\mathbb{R}}p(x)^q\:\mathrm{d}x} = \beta_q^2,
\end{equation}
which are known as $q$-mean and $q$-variance, respectively. 
These are generalizations of standard first and second moments, and tend to the usual mean 
and variance, respectively, as {$q\to1$}.
The Tsallis entropy maximizer under the above constraints~\eqref{q-expect-defn}
is the univariate $q$-Gaussian distribution that has the form
\begin{equation}
\label{Gq1:formula}
p(x) = {\frac{1}{{\beta_q}K_{q,1}}}\left(1-{\frac{(1-q)}{(3-q)\beta_q^2}(x-\mu_q)^2}\right)_+^{\frac{1}{1-q}}
  \quad\text{for all } x\in\mathbb{R},
\end{equation}
where, {$y_+=\max(y,0)$} is called the Tsallis cut-off condition, 
which ensures that the above expression is well-defined, 
and {$K_{q,1}$} is the normalizing constant for the univariate distribution.
The function defined in~\eqref{Gq1:formula} is not integrable for {$q\geqslant3$}, and 
hence, $q$-Gaussian is a probability density function only when {$q<3$}. 

A multivariate form of the $q$-Gaussian distribution has been discussed in~\cite{Vignat_2007_jour_PhyA}. 
As with the case of Gaussian smoothing kernel, here also
we are only interested in the case when the $N$-dimensional $q$-mean is zero,
and the $q$-covariance matrix (generalization of the usual covariance matrix that follows from~\eqref{q-expect-defn})
is $\beta^2 I_{N\times N}$.
Then the $N$-variate $q$-Gaussian distribution can be expressed as 
\begin{equation}
\label{Gq:formula_q}
G_{q,\beta}(\eta) = {\frac{1}{K_{q,N} \beta^{N}}}\left(1-{\frac{(1-q)}{(N+2-Nq)}\frac{\Vert\eta\Vert^2}{\beta^2}}
\right)_+^{\frac{1}{1-q}}
\end{equation}
for all {$\eta\in\mathbb{R}^N$}, where the normalizing constant is
\begin{equation}
K_{q,N} = \left\{\begin{array}{lcl}
		\left(\frac{N+2-Nq}{1-q}\right)^{\frac{N}{2}}
		\frac{\pi^{N/2} \Gamma\left(\frac{2-q}{1-q}\right)}{\Gamma\left(\frac{2-q}{1-q}+\frac{N}{2}\right)} &\text{for}  &q<1,
		\\\\
		\left(\frac{N+2-Nq}{q-1}\right)^{\frac{N}{2}}
		\frac{\pi^{N/2} \Gamma\left(\frac{1}{q-1}-\frac{N}{2}\right)}{\Gamma\left(\frac{1}{q-1}\right)} &\text{for}  
		& 1<q<\left(1+\frac{2}{N}\right).
                \end{array}\right.
\label{normalizing_const}
\end{equation}
In this multi-dimensional case, the distribution is only defined for $q<1+\frac{2}{N}, q\neq1$,
with Gaussian distribution being obtained in the limit of $q\to1$.
However, its usual moments are finite over a smaller range of $q$'s. For instance,
the mean is defined (and is finite) only for $q<(1+\frac{2}{N+1})$, and variance terms are finite
for $q<(1+\frac{2}{N+2})$. Some useful facts about the moments can be inferred from
Proposition~\ref{moments}, presented in Section~\ref{GqSF_convergence}.
Apart from the special cases mentioned in Table~\ref{tab_kernel},
one can see that the distribution 
has one-one correspondence with the Students'-$t$ distribution
for {$q\in\big(1,1+\frac{2}{N}\big)$}, and $q=-1$ provides the semi-circle distribution.

In this paper, we study the multivariate $q$-Gaussian distribution as a smoothing kernel, 
and develop smoothed functional algorithms based on it.
In general, we will denote the kernel by $G_{q,\beta}$ as in~\eqref{Gq:formula_q}.
However, as in the Gaussian SF methods presented in~\cite{Bhatnagar_2007_jour_TOMACS},
we will often use the ``standard'' distribution, \textit{i.e.}, when $\beta=1$.
Unlike common notions of Gaussian distribution, here the standard case does not imply that the components 
are independent. However, Proposition~\ref{moments} will show that the components will be uncorrelated 
in this case. We will refer to the standard case as $G_q \equiv G_{q,1}$ for convenience.
One can verify that the support set of $G_q$ is given by
\begin{equation}
\Omega_q = \left\{\begin{array}{lcl}
		\left\{\eta\in\mathbb{R}^N: \Vert\eta\Vert^2 < \frac{N+2-Nq}{1-q}\right\}
		&\text{for}  &q<1,
		\\\\
		\mathbb{R}^N &\text{for} & 1<q<\left(1+\frac{2}{N}\right).
                \end{array}\right.
\label{support}
\end{equation}
For the case when we consider $G_{q,\beta}$ centered at $\mu$, we will simply use the notation 
$\mu+\beta\Omega_q$ to denote the support of the distribution.

\section{\lowercase{$q$}-Gaussian based Smoothed Functional Algorithms}
\label{qGrad}
\subsection{\lowercase{$q$}-Gaussian as a Smoothing Kernel}
\label{qGaussian_SF}

The first step in applying $q${-}Gaussians for SF algorithms is to ensure that 
the distribution satisfies the Rubinstein conditions (properties (P1)--(P5) in Section~\ref{preliminaries_SF}).
The rest of the paper uses the multivariate form of $q$-Gaussian, $G_{q,\beta}$,
given in~\eqref{Gq:formula_q}, and the corresponding smoothed functionals
(one and two-sided) are given by $S_{q,\beta}^i$, $i=1,2$.

\begin{proposition}
The $N$-dimensional $q$-Gaussian distribution, $G_{q,\beta}$ \eqref{Gq:formula_q}, with $q$-covariance {$\beta^2I_{N\times N}$}
satisfies the kernel properties (P1)--(P5) 
for all {$q<\left(1+\frac{2}{N}\right)$}, {$q\neq1$}. 
\end{proposition}
\proof
\begin{enumerate}[(P1)]
\item 
From~\eqref{Gq:formula_q}, it is evident that
{$G_{q,\beta}(\eta) = \displaystyle\frac{1}{{\beta}^N}G_q\left(\frac{\eta}{\beta}\right)$}.
\item 
For {$1<q<\left(1+\frac{2}{N}\right)$}, {$G_{q,\beta}(\eta)>0$} for all {$\eta\in\mathbb{R}^N$}. Thus,
\begin{equation}
\nabla_{\eta}G_{q,\beta}(\eta) = -{\frac{2\eta}{(N+2-Nq)\beta^2}} \frac{G_{q,\beta}(\eta)}
{\left(1-{\frac{(1-q)}{(N+2-Nq)\beta^2} \Vert\eta\Vert^2}\right)}.
\label{grad}
\end{equation}
For {$q<1$},~\eqref{grad} holds when {$\eta\in\beta\Omega_q$} (the support set).
On the other hand, whenever {$\eta\notin\beta\Omega_q$}, we have {$G_{q,\beta}(\eta) = 0$} and hence, {$\nabla_{\eta}G_{q,\beta}(\eta) = 0$}.
Thus, {$G_{q,\beta}(\eta)$} is differentiable for {$q>1$}, and piecewise differentiable for {$q<1$}.
\item
{$G_{q,\beta}(\eta)$} is a distribution for {$q<\left(1+\frac{2}{N}\right)$} and hence, the corresponding one-sided SF, {$S_{q,\beta}^1$}, can be written as
{$(S_{q,\beta}^1 f)(\theta) = \mathsf{E}_{G_{q,\beta}(\eta)}[f(\theta-\eta)]$}.
\item
{$G_{q,\beta}$} is a distribution satisfying
{$\lim\limits_{\beta\to0}G_{q,\beta}(0) = \infty$}. 
So, {$\lim\limits_{\beta\to0}G_{q,\beta}(\eta) = \delta(\eta)$}.
\item
This property trivially holds due to convergence in mean as
\begin{displaymath}
\lim_{\beta\to0}(S_{q,\beta}^1 f)(\theta) =\int_{\mathbb{R}^N}{\lim_{\beta\to0}G_{q,\beta}(\eta)f(\theta-\eta)d\eta}
=\int_{\mathbb{R}^N}{\delta(\eta)f(\theta-\eta)\:\mathrm{d}\eta}
=f(\theta). 
\end{displaymath}
\end{enumerate}
The above claims hold in a similar manner for $S_{q,\beta}^2$, and hence, the result.
\qed

From the above result, it follows that $q$-Gaussian can be used as a kernel function, and hence,
given a particular value {$q\in\big(-\infty,1\big)\bigcup\big(1,1+\frac{2}{N}\big)$} and some {$\beta>0$}, the one-sided and two-sided
SFs of any function {$f\in\mathbb{R}^C$}
for compact $C\subset\mathbb{R}^N$  are respectively given by
\begin{align}
\label{qG_SF1_defn}
(S_{q,\beta}^1 f)(\theta) 
&= \int\limits_{\beta\Omega_q} G_{q,\beta}(\eta) f(\theta-\eta) \:\mathrm{d}\eta
= \int\limits_{\theta+\beta\Omega_q} G_{q,\beta}(\theta-\eta) f(\eta) \:\mathrm{d}\eta,
\\(S_{q,\beta}^2 f)(\theta)
&= \frac{1}{2}\int\limits_{\beta\Omega_q}G_{q,\beta}(\eta)
\big(f(\theta+\eta)+f(\theta-\eta)\big)\:\mathrm{d}x,
\label{qG_SF2_defn}
\end{align}
where the nature and smoothing properties of the SFs are controlled by both $q$ and $\beta$.

\subsection{Gradient estimation via $q$-Gaussian SF}

The objective is to estimate the gradient of the average cost {$\nabla_{\theta}J(\theta)$} using the SF approach, where existence of 
{$\nabla_{\theta}J(\theta)$} follows from Assumption~\ref{differentiable}. 
For the one-sided SF, 
the gradient of the SF (smoothed gradient) 
can be obtained simply by considering the derivative of~\eqref{qG_SF1_defn},
where the change of the integral and derivative can be done by Leibnitz rule
since $G_{q,\beta}(\theta-\eta)$ is differentiable with respect to $\theta$.
For $q>1$, the support set $(\theta+\beta\Omega_q) \equiv \mathbb{R}^N$,
and hence, the one-sided smoothed gradient can be written as
\begin{equation}
\nabla_{\theta} (S_{q,\beta}^1 J)(\theta) = 
\int\limits_{\theta+\beta\Omega_q} \nabla_{\theta}G_{q,\beta}(\theta-\eta) J(\eta) \:\mathrm{d}\eta\;.
\label{eq_1SFgrad_step1}
\end{equation}
For $q<1$, we observe that the support set is a function of $\theta$. So an
additional integral term should be present due to Leibnitz rule, where the integration is 
over the surface of the set $(\theta+\beta\Omega_q)$. However, the integrand involves
the term $G_{q,\beta}({\theta-\eta})$, which is zero over this surface, and hence, the 
additional term turns out to be zero.

As there is no functional relationship between {$\theta$} and {$\eta$}, 
\textit{i.e.}, $\displaystyle\frac{\:\mathrm{d}\eta^{(j)}}{\:\mathrm{d}\theta^{(i)}} = 0$ 
for all $i,j$, the $i^{th}$ coordinate of the gradient of $G_{q,\beta}$ is expressed as
\begin{align}
\nabla_{\theta}^{(i)} G_{q,\beta}(\theta-\eta) 
&= \frac{1}{\beta^NK_{q,N}}\frac{2\left(\eta^{(i)}-\theta^{(i)}\right)}{\beta^2(N+2-Nq)}
\left(1- \frac{(1-q)\sum_{k=1}^{N}
\left(\theta^{(k)}-\eta^{(k)}\right)^2}{(N+2-Nq)\beta^2}\right)^{\frac{q}{1-q}} 
\nonumber\\
&= \frac{2}{\beta^2(N+2-Nq)}\frac{\left(\eta^{(i)}-\theta^{(i)}\right)}
{\rho{(\frac{\theta-\eta}{\beta})}} G_{q,\beta}(\theta-\eta)\;,
\label{grad_qG}
\end{align}
where {$\rho{(\eta)} = \left(1-{\frac{(1-q)}{(N+2-Nq)}{\Vert}\eta{\Vert}^2}\right)$}.
Hence, substituting {$\eta' = \frac{\eta-\theta}{\beta}$}, and using the symmetry of {$G_{q,\beta}(.)$} and {$\rho(.)$}, we can write
\begin{align}
\nabla_{\theta} (S_{q,\beta}^1 J)(\theta)
&= \left(\frac{2}{\beta(N+2-Nq)}\right)\int\limits_{\Omega_q} 
\frac{\eta'}{\rho{(\eta')}} G_{q}(\eta') J(\theta+\beta\eta') \:\mathrm{d}\eta'
\nonumber
\\&= \left(\frac{2}{\beta(N+2-Nq)}\right)\mathsf{E}_{G_{q}(\eta')}
\left[\left. \frac{\eta'}{\rho{(\eta')}} J(\theta+\beta\eta') \right| \theta \right]\;.
\label{grad_SF1_formula}
\end{align}
In the sequel (Proposition~\ref{grad_SF1_convergence}), we show that
{$\big\Vert \nabla_{\theta}(S_{q,\beta}^1 J)(\theta)-\nabla_{\theta}J(\theta)\big\Vert \to 0$} as {$\beta \to 0$}.
Hence, for large $M$ and small {$\beta$}, the form of gradient estimate suggested by~\eqref{grad_SF1_formula} is
\begin{equation}
\label{approx1}
\nabla_{\theta}J(\theta)\approx\left(\frac{2}{\beta(N+2-Nq)M}\right)
\sum_{n=0}^{M-1}\left(\frac{\eta_n J(\theta+\beta\eta_n)}{\rho{(\eta_n)}}\right)\;,
\end{equation}
where {$\eta_0,\eta_1,\ldots,\eta_{M-1}$} are independent identically distributed standard $q$-Gaussian distributed random vectors.
Considering that in two-timescale algorithms (discussed later),
the value of {$\theta$} is updated concurrently with the gradient estimation procedure,
we estimate {$\nabla_{\theta_n}J(\theta_n)$} at each stage.
By ergodicity assumption (Assumption~\ref{ergodic}), we can write~\eqref{approx1} as 
\begin{equation}
\label{estimate1}
\nabla_{\theta_n}J(\theta_n)\approx\left(\frac{2}{\beta ML(N+2-Nq)}\right) \sum_{n=0}^{M-1}\sum_{m=0}^{L-1}
\frac{\eta_n h(Y_m(\theta_n+\beta\eta_n))}{\left(1-\frac{(1-q)}{(N+2-Nq)}
\Vert\eta_n\Vert^2\right)}
\end{equation}
for large $L$, where the process {$\{Y_m(\theta_n+\beta\eta_n)\}$} has the same transition kernel as defined in Assumption~\ref{ergodic},
except that it is governed by parameter {$(\theta_n+\beta\eta_n)$}.

In a similar manner, based on~\eqref{SF2}, the gradient of the two-sided SF can be written as
\begin{equation}
\nabla_{\theta} (S_{q,\beta}^2 J)(\theta)
= \frac{1}{2}\int\limits_{\theta+\beta\Omega_q} \nabla_{\theta}G_{q,\beta}(\theta-\eta)J(\eta)\:\mathrm{d}\eta
+ \frac{1}{2}\int\limits_{\theta+\beta\Omega_q} \nabla_{\theta}G_{q,\beta}(\eta-\theta)J(\eta)\:\mathrm{d}\eta.
\end{equation}
From this, we can obtain the gradient as a conditional expectation as
\begin{equation}
\nabla_{\theta} (S_{q,\beta}^2 J)(\theta)
= \left(\frac{1}{\beta(N+2-Nq)}\right)\mathsf{E}_{G_{q}(\eta)}
\left[\left.\frac{\eta}{\rho{(\eta)}} \Big(J(\theta+\beta\eta)-J(\theta-\beta\eta)\Big)\right|\theta\right].
\label{grad_SF2_formula}
\end{equation}
In the sequel (Proposition~\ref{grad_SF2_convergence}) we show that
{$\big\Vert \nabla_{\theta}(S_{q,\beta}^2 J)(\theta)-\nabla_{\theta}J(\theta)\big\Vert \to 0$} as {$\beta \to 0$}, 
which can be used to approximate~\eqref{grad_SF2_formula}, for large $M$, $L$ and small {$\beta$}, as
\begin{equation}
\label{estimate2}
\nabla_{\theta_n}J(\theta_n)\approx\frac{1}{\beta ML(N+2-Nq)} \sum_{n=0}^{M-1}\sum_{m=0}^{L-1}
\frac{\eta_n\big(h(Y_m^1(\theta_n+\beta\eta_n)-h(Y_m^2(\theta_n-\beta\eta_n))\big)}{\left(1-\frac{(1-q)}{(N+2-Nq)}\Vert\eta_n\Vert^2\right)}
\end{equation}
where two processes $Y_m^1(\cdot)$ and $Y_m^2(\cdot)$ are respectively 
controlled by the parameters $(\theta_n+\beta\eta_n)$ and $(\theta_n-\beta\eta_n)$.

\subsection{Proposed Gradient Descent Algorithms}
\label{GqSF_algos}

We propose two-timescale algorithms based on the estimates obtained in~\eqref{estimate1} and \eqref{estimate2}.
Let {$(a_n)_{n\geqslant0}$} and {$(b_n)_{n\geqslant0}$} be two step-size sequences satisfying the following.
\begin{assumption}
\label{stepsize}
{$(a_n)_{n\geqslant0}$}, {$(b_n)_{n\geqslant0}$} are positive sequences 
satisfying the following:
{$\displaystyle\sum_{n=0}^{\infty}a_n^2 <\infty$}, {$\displaystyle\sum_{n=0}^{\infty}b_n^2 <\infty$},
{$\displaystyle\sum_{n=0}^{\infty}a_n = \sum_{n=0}^{\infty}b_n = \infty$} and {$a_n = o(b_n)$}, 
\textit{i.e.}, {$\frac{a_n}{b_n}\to0$} as {$n\to\infty$}.
\end{assumption}

It must be noted that in the algorithms, although $M$ is chosen to be a large quantity
(to ensure convergence), the quantity $L$ is arbitrarily picked and can be any finite positive number.
The averaging of the inner summation in~\eqref{estimate1} and~\eqref{estimate2} is obtained in our
algorithms using two-timescale stochastic approximation. In principle, one may select {$L=1$}. 
However, it is generally observed that a value of $L$ typically between $5$ and $500$ results in better 
performance~\cite{Bhatnagar_2007_jour_TOMACS}. Further, the algorithms require generation of $N$-dimensional random vectors, 
consisting of uncorrelated $q$-Gaussian distributed random variates.
This method is described in the appendix.

For {$\theta \in\mathbb{R}^N$}, let {$\mathcal{P}_C(\theta)$}
represent the projection of {$\theta$} onto the set $C$. 
For simulation, we need to project the perturbed random vectors {$(\theta_n+\beta\eta_n)$} onto $C$
using this projection. However, for convenience, we will refer to the 
process as $\{Y_m(\theta_n+\beta\eta_n)\}$ without explicitly mentioning the projection.
{$(Z_n)_{n\geqslant0}$} are $N$-dimensional vectors used to estimate 
{$\nabla_{\theta}J(\theta)$} in the  recursions.
Note that the term within brackets in Step 7 is a scalar.

\begin{algorithm}[ht]
\caption{The G$q$-SF1 Algorithm}
\SetAlgoNoLine
Fix $M$, $L$, $q$ and $\beta$\;
Set {$Z_0 = 0 \in\mathbb{R}^N$}, and fix the initial parameter vector {$\theta_0 \in C$}\;
\For{{$n=0$} to {$M-1$}}{
Generate a random vector {$\eta_n$} from a standard $N$-dimensional $q$-Gaussian distribution\;
\For{{$m=0$} to {$L-1$}}{
Generate the simulation {$Y_{nL+m}(\theta_n+\beta\eta_n)$} governed with parameter 
{$\mathcal{P}_C(\theta_n+\beta\eta_n)$}\;
Gradient estimate {$Z_{nL+m+1} = (1-b_n)Z_{nL+m}
+b_n\eta_n\left[\frac{2h(Y_{nL+m}(\theta_n+\beta\eta_n))}
{\beta(N+2-Nq)\left(1-\frac{(1-q)}{(N+2-Nq)}\Vert\eta_n\Vert^2\right)}\right]$}\; 
}
Update parameter vector
{$\theta_{n+1} = \mathcal{P}_C\left(\theta_n - a_n Z_{nL}\right)$}\;
}
Output {$\theta_M$} as the final parameter vector\;
\end{algorithm}

The G$q$-SF2 algorithm is similar to the G$q$-SF1 algorithm,
except that we use two parallel simulations {$Y_{nL+m}^1(\theta_n+\beta\eta_n)$} and {$Y_{nL+m}^2(\theta_n-\beta\eta_n)$}, 
and update the gradient estimate using the single-stage cost function of both simulations as in~\eqref{estimate2}.
We note that during update of the gradient estimate (Steps 5--8), the two simulations $Y_m^1(\cdot)$ and $Y_m^2(\cdot)$
are not affected by each other.

\begin{algorithm}[h]
\caption{The G$q$-SF2 Algorithm}
\SetAlgoNoLine
Fix $M$, $L$, $q$ and $\beta$\;
Set {$Z_0 = 0 \in\mathbb{R}^N$}, and fix the initial parameter vector {$\theta_0 \in C$}\;
\For{{$n=0$} to {$M-1$}}{
Generate a random vector {$\eta_n$} from a standard $N$-dimensional $q$-Gaussian distribution\;
\For{{$m=0$} to {$L-1$}}{
Generate two simulations {$Y_{nL+m}^1(\theta_n+\beta\eta_n)$} and {$Y_{nL+m}^2(\theta_n-\beta\eta_n)$} 
governed with parameters {$\mathcal{P}_C(\theta_n+\beta\eta_n)$} and {$\mathcal{P}_C(\theta_n-\beta\eta_n)$}, respectively\;
Update {$Z_{nL+m+1} = (1-b_n)Z_{nL+m} +b_n\eta_n\left[\frac{h(Y_{nL+m}^1(\theta_n+\beta\eta_n))-h(Y_{nL+m}^2(\theta_n-\beta\eta_n))}
{\beta(N+2-Nq)\left(1-\frac{(1-q)}{(N+2-Nq)}\Vert\eta_n\Vert^2\right)}\right]$}\; 
}
Update parameter vector
{$\theta_{n+1} = \mathcal{P}_C\left(\theta_n - a_n Z_{nL}\right)$}\;
}
Output {$\theta_M$} as the final parameter vector\;
\end{algorithm}

\section{Convergence of the proposed Algorithms}
\label{GqSF_convergence}

We prove that the algorithms converge to stationary points of an associated ODE.
The techniques we use are more straightforward as compared to~\cite{Bhatnagar_2007_jour_TOMACS}.
Before presenting the details of convergence analysis,
we present the following result on $q$-Gaussians.
It provides an expression for the moments of
$N$-variate $q$-Gaussian distributed random vector. This 
is a consequence of the results presented in~\cite{Gradshteyn_1994_book_Elesevier}.
This result plays a key role in the proofs discussed below.

\begin{proposition}
\label{moments}
Suppose {$X = \left(X^{(1)}, X^{(2)}, \ldots, X^{(N)}\right)$} is a $N$-dimensional
random vector, where the components are uncorrelated and identically distributed, each being distributed according to a $q$-Gaussian
distribution with zero $q$-mean and unit $q$-variance, with parameter {$q\in \big(-\infty,1\big)\bigcup\big(1,1+\frac{2}{N}\big)$}.
Also, let {$\rho(X) = \left(1 - \frac{(1-q)}{(N+2-Nq)}\Vert{X}\Vert^2\right)$}.
Then, for any {$k, k_1, k_2, \ldots, k_N \in \mathbb{Z}^{+}\bigcup\{0\}$}, we have
\begin{equation}
\mathsf{E}_{G_q} \left[\frac{\left(X^{(1)}\right)^{k_1}\left(X^{(2)}\right)^{k_2}\ldots\left(X^{(N)}\right)^{k_N}}{\left(\rho(X)\right)^k}\right]
= \left\{\begin{array}{l} 
\bar{K}\displaystyle\left(\frac{N+2-Nq}{|1-q|}\right)^{\sum\limits_{i=1}^{N}\frac{k_i}{2}} 
\left(\prod\limits_{i=1}^{N}\frac{k_i!}{2^{k_i}\left(\frac{k_i}{2}\right)!}\right)
\\ \hspace{10mm}\text{if } k_i \text{ is even for all } i = 1,2,\ldots,N,
\\\\
0  
\hspace{8mm}\text{otherwise, }
\end{array}\right.
\label{moments_eqn}
\end{equation}
where
\begin{equation}
\\\bar{K}	= \left\{\begin{array}{lcl}
		\frac{\Gamma\left(\frac{1}{1-q}-k+1\right) \Gamma\left(\frac{1}{1-q}+1+\frac{N}{2}\right)}
		{\Gamma\left(\frac{1}{1-q}+1\right) \Gamma\left(\frac{1}{1-q}-k+1+\frac{N}{2}+ \sum\limits_{i=1}^{N}\frac{k_i}{2}  \right)}
		&\text{if} 
		&q\in(-\infty,1),
		\\\\
		\frac{\Gamma\left(\frac{1}{q-1}\right) \Gamma\left(\frac{1}{q-1}+k-\frac{N}{2} -\sum\limits_{i=1}^{N}\frac{k_i}{2}\right)}
		{\Gamma\left(\frac{1}{q-1}+k\right) \Gamma\left(\frac{1}{q-1}-\frac{N}{2}\right)}
		&\text{if} 
		&q\in\left(1,1+\frac{2}{N}\right),
                \end{array}\right.
\end{equation}
exists only if the arguments in the above Gamma functions are positive, which holds for
{$k<\left(1+\frac{1}{1-q}\right)$} if {$q<1$}, and 
{$\left(\frac{1}{q-1}-\frac{N}{2}\right)>\left(\sum_{i=1}^{N}\frac{k_i}{2}-k\right)$} if {$1<q<\left(1+\frac{2}{N}\right)$}.
\end{proposition}

\proof
Since {$\Sigma_q=I_{N\times N}$}, and {$\rho(X)$} is non-negative over {$\Omega_q$}, we have
\begin{align*}
&\mathsf{E}_{G_q(X)} \left[\frac{\left(X^{(1)}\right)^{k_1}\left(X^{(2)}\right)^{k_2}\ldots\left(X^{(N)}\right)^{k_N}}{\left(\rho(X)\right)^k}\right]
\\&= \frac{1}{K_{q,N}} \int\limits_{\Omega_q} \left(x^{(1)}\right)^{k_1}\left(x^{(2)}\right)^{k_2}\ldots\left(x^{(N)}\right)^{k_N} 
\left(1 - \frac{(1-q)}{\big(N+2-Nq\big)}\Vert{x}\Vert^2\right)^{\frac{1}{1-q}-k}\:\mathrm{d}x. 
\end{align*}

The second equality in~\eqref{moments_eqn} can be easily proved. If for some {$i = 1,\ldots,N$}, {$k_i$} is odd, then the above function is odd, 
and its integration is zero over {$\Omega_q$}, which is symmetric with respect to any axis by definition.
For the other cases, since the function is even, the integral is same over every orthant. Hence, we may consider
the integration over the first orthant, \textit{i.e.}, where each component is positive. 
For {$q<1$}, we can reduce the above integral, using~\cite[equation (4.635)]{Gradshteyn_1994_book_Elesevier}, to obtain
\begin{align}
\mathsf{E}_{G_q(X)} &\left[\frac{\prod_{i=1}^{N}\left(X^{(i)}\right)^{k_i}}{\left(\rho(X)\right)^k}\right]
&= 
\frac{\prod\limits_{i=1}^{N} \Gamma\left(\frac{k_i+1}{2}\right)}{K_{q,N}\Gamma( \bar{k} )}
\left(\frac{N+2-Nq}{1-q}\right)^{\bar{k}} 
\int\limits_{0}^{1} (1-y)^{\left(\frac{1}{1-q}-k\right)}y^{\left(\bar{k}-1\right)}\mathrm{d}x 
\label{integral}
\end{align}
where we set {$\bar{k} = \left(\frac{N}{2}+\sum_{i=1}^{N}\frac{k_i}{2}\right)$}.
One can observe that the integral in~\eqref{integral} is in the form of a Beta function.
Since {$k_i$}'s are even,
we can expand {$\Gamma\left(\frac{k_i+1}{2}\right)$} using the expansion of Gamma function of half-integers to get
{$\Gamma\left(\frac{k_i+1}{2}\right) = \frac{k_i!}{2^{k_i}\left(\frac{k_i}{2}\right)!}\sqrt{\pi}$}.
The claim can be obtained by substituting {$K_{q,N}$} from~\eqref{normalizing_const} and using the 
relation {$B(m,n) = \frac{\Gamma(m)\Gamma(n)}{\Gamma(m+n)}$}. It is easy to verify that all the Gamma functions in the equality
are positive provided {$k<\left(1+\frac{1}{1-q}\right)$}.
The result for the interval {$1<q<\left(1+\frac{2}{N}\right)$} can be proved in a similar way
(see equations (4.635) and (4.636) of~\cite{Gradshteyn_1994_book_Elesevier}). 
However, in this case the Gamma functions are positive if {$k, k_1, k_2, \ldots, k_N\in\mathbb{Z}^{+}\bigcup\{0\}$}
satisfy the mentioned condition. It may be noted here that this is always true for any dimension if {$k>\sum_{i=1}^{N}\frac{k_i}{2}$}
since $q$-Gaussians are defined only when {$\frac{1}{q-1}>\frac{N}{2}$}.
\qed

It is easy to verify the following result in the limiting case of $q\to1$ as 
\begin{displaymath}
\lim_{q\to1}\mathsf{E}_{G_q(X)} 
\left[\frac{\left(X^{(1)}\right)^{k_1}\left(X^{(2)}\right)^{k_2}\ldots\left(X^{(N)}\right)^{k_N}}{\left(\rho(X)\right)^k}\right]
= \prod_{i=1}^{N} \mathsf{E}_{G(X)} \left[\left(X^{(i)}\right)^{k_i}\right]\;.
\end{displaymath}
This ensures that the subsequent convergence analysis also holds for Gaussian SF.

\subsection{Convergence of G$q$-SF1 Algorithm}
First, let us consider the update of the gradient
along the faster timescale, \textit{i.e.}, Step~7 of the G$q$-SF1 algorithm.
We rewrite this as the update iteration
\begin{equation}
\label{faststep}
Z_{p+1} = Z_p + \tilde{b}_p\big[ g(Y_p) - Z(p) \big]
\qquad\qquad \text{ for all } p\geqslant0,
\end{equation}
where we use $g(Y_p)$ to denote
\begin{displaymath}
 g(Y_p) = \left(\frac{2h(Y_p(\tilde{\theta}_p+\beta\tilde{\eta}_p))}{\beta(N+2-Nq) \rho{(\tilde{\eta}_p)}}\right) \tilde{\eta}_p\;.
\end{displaymath}
The purpose of rewriting the update as in~\eqref{faststep} is to emphasize
that the gradient estimation at each stage is not carried out in a disjoint manner;
rather the previous estimate is updated at each epoch. 
Here, for each  {$n\geqslant0$} and {$nL\leqslant p<(n+1)L$}, 
we use the notation {$\tilde{\theta}_p = \theta_n$}, {$\tilde{\eta}_p = \eta_n$} and
{$\tilde{b}_p = b_n$}. It follows from Assumption~\ref{stepsize}
that {$a_p = o(\tilde{b}_p)$}, {$\sum_p \tilde{b}_p = \infty$} and {$\sum_p \tilde{b}_p^2 < \infty$}. 
We also note that {$\rho{(.)}$} is defined as in~\eqref{grad_qG}, and
{$\{Y_p(\tilde{\theta}_p+\beta\tilde{\eta}_p):p\in\mathbb{N}\}$} is a Markov process parameterized 
by {$\mathcal{P}_C{(\tilde{\theta}_p+\beta\tilde{\eta}_p)}$}.
Now, let {$\mathcal{G}_p = \sigma\big(\tilde{\theta}_k, \tilde{\eta}_k, Y_k(\tilde{\theta}_k+\tilde{\eta}_k), k\leqslant p\big)$} 
denote the $\sigma$-field generated by the mentioned quantities. We can observe that {$(\mathcal{G}_p)_{p\geqslant0}$} is a filtration,
where {$g(Y_p)$} is {$\mathcal{G}_p$}-measurable.

We summarize the results presented in~\cite[Chapter 6, Lemma 3 -- Theorem 9]{Borkar_2008_book_Cambridge} 
in the following theorem. This result leads to the 
stability and convergence of iteration~\eqref{faststep}, which runs on the faster timescale. 

\begin{theorem}
\label{borkar_cor_8}
Consider the iteration, {$x_{p+1} = x_p + \gamma_p\big(f(x_p,Y_p)+M_p\big)$}.
Let the following conditions hold:
\begin{enumerate}
\item 
{$\{Y_p:p\in\mathbb{N}\}$} is a Markov process satisfying Assumptions~\ref{ergodic} and~\ref{lyapunov},
\item
for each {$x\in\mathbb{R}^N$} and {$x_p\equiv x$} for all $p\in\mathbb{N}$,
{$Y_p$} has a unique invariant probability measure {$\nu_{x}$},
\item
{$(\gamma_p)_{p\geqslant0}$} are step-sizes satisfying {$\sum\limits_{p=0}^{\infty} \gamma_p=\infty$} 
and {$\sum\limits_{p=0}^{\infty} \gamma_p^2<\infty$},
\item
{$f(.,.)$} is Lipschitz continuous in its first argument uniformly w.r.t the second, 
\item
{$M_p$} is a martingale difference noise term with bounded variance,
\item
if {$\tilde{f}\big(x,\nu_x\big) = \mathsf{E}_{\nu_x} \big[f(x,Y)\big]$}, then the limit
{$\hat{f}\big(x(t)\big) = \displaystyle\lim\limits_{a\uparrow\infty} \frac{\tilde{f}\big(ax(t),\nu_{ax(t)}\big)}{a}$}
exists uniformly on compacts, and
\item
the ODE {$\dot{x}(t) = \hat{f}\big(x(t)\big)$} is well-posed
and has the origin as the unique globally asymptotically stable equilibrium.
\end{enumerate}
Then the update {$x_p$} satisfies {$\sup_p \Vert{x_p}\Vert < \infty$}, almost surely,
and converges to the stable fixed points of the ordinary differential equation (ODE)
\begin{displaymath}
\dot{x}(t) = \tilde{f}\big(x(t),\nu_{x(t)}\big).
\end{displaymath}
\end{theorem}

We rewrite the update~\eqref{faststep} as
\begin{equation}
Z_{p+1} = Z_p + \tilde{b}_p\big( E[g(Y_p)|\mathcal{G}_{p-1}] - Z_p + A_p\big),
\label{faststep1}
\end{equation}
where {$A_p = g(Y_p) - E[g(Y_p)|\mathcal{G}_{p-1}]$} is {$\mathcal{G}_p$}-measurable. 
The following result shows that {$(A_p,\mathcal{G}_p)_{p\geqslant0}$} satisfies 
Condition~5 in Theorem~\ref{borkar_cor_8}.

\begin{lemma}
\label{A_n_convergence}
For all values of {$q\in\big(-\infty,1\big)\bigcup\big(1,1+\frac{2}{N}\big)$},
{$(A_p,\mathcal{G}_p)_{p\in\mathbb{N}}$} is a martingale difference sequence with a bounded variance. 
\end{lemma}
\proof
For this proof, we will simply write $Y_p$ to denote the process 
parameterized $\mathcal{P}_C(\tilde{\theta}_p+\beta\tilde{\eta}_p)$,
as considered in~\eqref{faststep}.
It is easy to see that {$\mathsf{E}[A_{p}|\mathcal{G}_{p-1}] = 0$} for all $p\geqslant0$.
So {$(A_p,\mathcal{G}_p)_{p\in\mathbb{N}}$} is a martingale difference sequence. 
For computing the variance, we expand the terms as
\begin{align*}
&\mathsf{E}\left[\left.\Vert A_{p}\Vert^2\right|\mathcal{G}_{p-1}\right]
\\&\leqslant \frac{8}{\beta^2(N+2-Nq)^2} \mathsf{E}\left[\left.
\left(\frac{\Vert\tilde{\eta}_p\Vert h(Y_p)}{\rho{(\tilde{\eta}_p)}}\right)^2 +
\left(\left.\mathsf{E}\left[\frac{\Vert\tilde{\eta}_p\Vert h(Y_p)}{\rho{(\tilde{\eta}_p)}}\right|\mathcal{G}_{p-1}\right]\right)^2
\right|\mathcal{G}_{p-1}\right].
\end{align*}
Applying conditional Jensen's inequality on the second term, we obtain
\begin{align}
\mathsf{E}\left[\left.\Vert A_{p}\Vert^2\right|\mathcal{G}_{p-1}\right]
&\leqslant \frac{16}{\beta^2(N+2-Nq)^2} 
\mathsf{E}\left[\left.\frac{\Vert\tilde{\eta}_p\Vert^2}{{\rho{(\tilde{\eta}_p)}^2}}{h^2(Y_p)}\right| \mathcal{G}_{p-1}\right].
\label{Jensen_G1}
\end{align}
For {$q\in(-\infty,1)$}, 
we use Holder's inequality to write~\eqref{Jensen_G1} as
\begin{align*}
\mathsf{E}\left[\left.\Vert A_{p}\Vert^2\right|\mathcal{G}_{p-1}\right]
&\leqslant \frac{16}{\beta^2(N+2-Nq)^2} 
\sup\limits_{\eta} \left(\frac{\Vert\tilde{\eta}_p\Vert^2}{{\rho{(\tilde{\eta}_p)}^2}}\right)
\mathsf{E}\left[\left.{h^2(Y_p)}\right| \mathcal{G}_{p-1}\right]
\\&\leqslant \frac{16}{\beta^2 (1-q)(N+2-Nq)} \mathsf{E}\left[\left.{h^2(Y_p)}\right| \mathcal{G}_{p-1}\right],
\end{align*}
since, {$\Vert\eta\Vert^2 < \frac{N+2-Nq}{1-q}$} and {$\rho(\eta) \geqslant1$} for all {$\eta\in\Omega_q$}.
By Lipschitz continuity of $h$, there exists {$\alpha_1>0$} such that 
{$\vert h(Y_p) \vert \leqslant \alpha_1(1+\Vert{Y_p}\Vert)$} for all $p$,
and hence, by Assumption~\ref{lyapunov}, we can claim 
\begin{align}
\mathsf{E} \left[h(Y_p)^2|\mathcal{G}_{p-1}\right]
\leqslant 2\alpha_1^2\left( 1+ \mathsf{E} \left[\Vert{Y_p}\Vert^2|\mathcal{G}_{p-1}\right]\right)
<\infty \text{ a.s.}
\label{eq_Jensen_G1}
\end{align}
On the other hand,
for {$q\in\big(1,1+\frac{2}{N}\big)$}, we apply Cauchy-Schwartz inequality for
each of the components in~\eqref{Jensen_G1} to obtain
\begin{align*}
\mathsf{E}\left[\left.\Vert A_{p}\Vert^2\right|\mathcal{G}_{p-1}\right]
&\leqslant \frac{16}{\beta^2(N+2-Nq)^2} \sum_{j=1}^{N}
\mathsf{E}\left[\left.\frac{\left(\tilde{\eta}_p^{(j)}\right)^2}{{\rho{(\tilde{\eta}_p)}^2}}{h^2(Y_p)}\right| \mathcal{G}_{p-1}\right]
\\&\leqslant \frac{16}{\beta^2(N+2-Nq)^2} \sum_{j=1}^{N}
\mathsf{E}\left[\frac{\left(\tilde{\eta}_p^{(j)}\right)^4}{{\rho{(\tilde{\eta}_p)}^4}}\right]^{1/2}
\mathsf{E}\left[\left.{h^4(Y_p)}\right| \mathcal{G}_{p-1}\right]^{1/2}\;,
\end{align*}
where $\eta^{(j)}$ denotes the $j^{th}$ coordinate of $\eta$.
The second expectation can be shown to be finite as in~\eqref{eq_Jensen_G1}, while we
apply Proposition~\ref{moments} to study the existence of {$\mathsf{E}\left[\frac{\left(\eta^{(j)}\right)^4}{{\rho{(\eta)}^4}}\right]$}.
We can observe that in this case, {$k=4$} and {$k_i = 4$} if {$i=j$}, otherwise {$k_i=0$}, 
and so {$k>\sum_{i=1}^{N}\frac{k_i}{2}$}. Proposition~\ref{moments} ensures that the term is finite, 
and hence, the claim.
\qed

For the slower timescale, we write the parameter update (Step~9 of G$q$-SF1) as
\begin{displaymath}
\theta_{n+1} = \mathcal{P}_C\left(\theta_n - \tilde{b}_n\zeta_n\right),
\end{displaymath}
where {$\zeta_n=\frac{a_n}{\tilde{b}_n}Z_{nL}=o(1)$} since {$a_n = o(\tilde{b}_n)$}. 
Thus, the parameter update recursion can be seen to track the ODE 
\begin{equation}
 \dot{\theta}(t) = 0.
\label{fastode_theta}
\end{equation}
Hence, the recursion {$(\theta_n)_{n\geqslant0}$} appears quasi-static
when viewed from the timescale of {$(\tilde{b}_n)$}, and hence, in the update~\eqref{faststep1},
one may let {$\tilde{\theta}_p \equiv \theta$} and {$\tilde{\eta}_p \equiv \eta$} for all {$p\geqslant0$}.
Consider the following ODE
\begin{align}
\dot{Z}(t) &= \frac{2\eta J(\theta+\beta\eta)}{\beta(N+2-Nq)\rho{(\eta)}} - Z(t).
\label{fastode}
\end{align}

\begin{lemma}
\label{Z_bounded}
The sequence {$(Z_p)$} is uniformly bounded with probability 1. Further,
\begin{displaymath}
\left\Vert Z_p - \left(\displaystyle\frac{2\tilde{\eta}_p J(\tilde{\theta}_p+\beta\tilde{\eta}_p)}
{\beta(N+2-Nq)\rho{(\tilde{\eta}_p)}}\right) \right\Vert\to 0 
\end{displaymath}
almost surely as {$p\to\infty$}.
\end{lemma}
\proof
It can be easily verified that iteration~\eqref{faststep1} satisfies all the conditions of Theorem~\ref{borkar_cor_8}.
Thus, by Theorem~\ref{borkar_cor_8}, {$(Z_p)_{p\geqslant0}$} converges to the stable point of ODE~\eqref{fastode} as 
\begin{displaymath}
\mathsf{E}_{\nu_{(\theta+\beta\eta)}}\left[\frac{2\eta\, h(Y_p)}{\beta(N+2-Nq) \rho{(\eta)}} \right]
= \frac{2\eta J(\theta+\beta\eta)}{\beta(N+2-Nq)\rho{(\eta)}}\;.
\end{displaymath}
We can also see that
\begin{displaymath}
\lim\limits_{a\uparrow\infty}
\frac{1}{a}\left(\frac{2\eta J(\theta+\beta\eta)}{\beta(N+2-Nq)\rho{(\eta)}} - aZ(t)\right) = -Z(t).
\end{displaymath}
All the conditions in Theorem~\ref{borkar_cor_8} are seen to be verified and the claim follows.
\qed

From Lemma~\ref{Z_bounded}, Step 9 of G$q$-SF1 can be rewritten as
\begin{align}
\theta_{n+1}
&= \mathcal{P}_C \left(\theta_n - a_n\left[
\frac{2\eta_n J(\theta_n+\beta\eta_n)}{ 
\beta(N+2-Nq)\rho{(\eta_n)}}\right]\right)	\nonumber
\\&= \mathcal{P}_C \bigg(\theta_n + a_n\left[-
\nabla_{\theta_n} J(\theta_n) + \Delta (\theta_n) + \xi_n\right]\bigg),
\label{slowstep}
\end{align}
where the error in the gradient estimate is given by
\begin{equation}
\Delta (\theta_n) = \nabla_{\theta_n} J(\theta_n)
- \nabla_{\theta_n} (S_{q,\beta}^1 J)(\theta_n)
\end{equation}
and the noise term is
\begin{align}
\xi_n &= \nabla_{\theta_n} (S_{q,\beta}^1 J)(\theta_n)
- \frac{2\eta_n J(\theta_n+\beta\eta_n)}{\beta(N+2-Nq)\rho{(\eta_n)}}			\nonumber
\\&= \frac{2}{\beta(N+2-Nq)}\bigg(
\mathsf{E}_{G_q(\eta)}\left[\left.\frac{\eta_n}{\rho{(\eta_n)}}J(\theta_n+\beta\eta_n)\right| \theta_n\right]
- \frac{\eta_n}{\rho{(\eta_n)}}J(\theta_n+\beta\eta_n)\bigg),
\label{noise_term}
\end{align}
which is a martingale difference term.
Let {$\mathcal{F}_n = \sigma(\theta_0,\ldots,\theta_n, \eta_0,\ldots,\eta_{n-1})$} denote the $\sigma$-field
generated by the mentioned quantities. We can observe that {$(\mathcal{F}_n)_{n\geqslant0}$} is a filtration,
where {$\xi_0,\ldots,\xi_{n-1}$} are {$\mathcal{F}_n$}-measurable for each {$n\geqslant0$}. 

We state the following result due to~\citeN[Theorem~5.3.1, pp 189--196]{Kushner_1978_book_Springer}, adapted to our scenario,
which leads to the convergence of the updates in~\eqref{slowstep}.
\begin{lemma}
\label{kushner_thm_5.3.1}
Given the iteration, {$x_{n+1} = \mathcal{P}_C \big(x_n + \gamma_n(f(x_n) + \xi_n)\big)$}, where
\begin{enumerate}
\item
{$\mathcal{P}_C$} represents a projection operator onto a closed and bounded constraint set {$C$}, 
\item 
{$f(.)$} is a continuous function,
\item
{$(\gamma_n)_{n\geqslant0}$} is a positive sequence satisfying {$\gamma_n\downarrow0$}, {$\sum_{n=0}^{\infty} \gamma_n=\infty$}, and 
\item
{$\sum_{n=0}^m \gamma_n\xi_n$} converges a.s.
\end{enumerate}
Under the above conditions, the update {$(x_n)$} converges almost surely 
to the set of asymptotically stable fixed points of the ODE
\begin{equation}
\label{projectedode}
\dot{x}(t) = \tilde{\mathcal{P}}_C \big(x(t),f(x(t))\big),
\end{equation}
where {$\tilde{\mathcal{P}}_C (x,y) = 
\lim_{\epsilon\downarrow0}\left(\frac{\mathcal{P}_C (x+\epsilon y)-x}{\epsilon}\right)$}
for all $x,y\in\mathbb{R}^N$.
\end{lemma}
A note on the above definition of $\tilde{P}_C$
is necessary. Observe that we may set $y=f(x)$
for any function $f:\mathbb{R}^N\mapsto\mathbb{R}^N$,
and if $x$ 
lies in the interior of $C$, 
then $\tilde{\mathcal{P}}_C (x,f(x)) = f(x)$, 
On the other hand, if $x$ 
lies on the boundary of $C$ 
and $(x+\epsilon f(x)) \notin C$ 
for any small $\epsilon>0$, 
then $\mathcal{P}_C (x,f(x))$ 
is the projection of $f(x)$
onto $C$,
which is unique as $C$ 
is convex.
The next result shows that the noise term $\xi_n$
satisfies the last condition in Lemma~\ref{kushner_thm_5.3.1},
while the subsequent result proves the error $\Delta(\theta_n)$ is considerably small.  
\begin{lemma}
\label{xi_n_convergence}
Let {$M_n = \sum_{k=0}^{n-1} a_k\xi_k$}. Then, for all values of {$q\in\big(-\infty,1\big)\bigcup\big(1,1+\frac{2}{N}\big)$},
{$(M_n,\mathcal{F}_n)_{n\in\mathbb{N}}$} is an almost surely convergent martingale sequence. 
\end{lemma}
\proof
We can easily observe that for all $k\geqslant0$, 
\begin{align*}
\mathsf{E}[\xi_{k}|\mathcal{F}_{k}]
=\frac{2}{\beta(N+2-Nq)}\bigg(
\mathsf{E}\left[\left.\frac{\eta_k J(\theta_k+\beta\eta_k)}{\beta\rho{(\eta_k)}}\right|\theta_k\right]
- \mathsf{E}\left[\left.\frac{\eta_k J(\theta_k+\beta\eta_k)}{\beta\rho{(\eta_k)}}\right|\mathcal{F}_k\right]
\bigg). 
\end{align*}
So {$\mathsf{E}[\xi_{k}|\mathcal{F}_{k}] = 0$},
since {$\theta_k$} is {$\mathcal{F}_{k}$}-measurable, whereas {$\eta_k$} is independent of {$\mathcal{F}_{k}$}.
It follows that {$(\xi_n,\mathcal{F}_n)_{n\in\mathbb{N}}$} is a martingale difference sequence, 
and hence {$(M_n,\mathcal{F}_n)_{n\in\mathbb{N}}$} is a martingale sequence. 
Now, use of conditional Jensen's inequality leads to
\begin{align*}
\mathsf{E}\left[\left.\Vert\xi_{k}\Vert^2\right|\mathcal{F}_{k}\right]
&= \sum_{j=1}^N \mathsf{E}\left[\left.\left(\xi_{k}^{(j)}\right)^2\right|\mathcal{F}_{k}\right]
\\&\leqslant \frac{16}{\beta^2(N+2-Nq)^2} \sum_{j=1}^N
\mathsf{E}\left[\left.\frac{\left(\eta_k^{(j)}\right)^2}{{\rho{(\eta_k)}^2}}{J(\theta_k+\beta\eta_k)^2}\right|\theta_k\right].
\end{align*}

\noindent
For any {$\eta\in\mathbb{R}^N$}, by definition
{$J\big(\theta_k+\beta\eta\big) = \mathsf{E} [h(Y_p)]$},
where the expectation is with respect to the stationary measure.
By Jensen's inequality, we can claim
{$J\big(\theta_k+\beta\eta\big)^2 \leqslant \mathsf{E} \left[h(Y_p)^2\right]$}
and {$J\big(\theta_k+\beta\eta\big)^4 \leqslant \mathsf{E} \left[h(Y_p)^4\right]$}
for all {$\eta\in\mathbb{R}^N$}.
Using these facts along with arguments similar to Lemma~\ref{A_n_convergence},
it can be seen that 
{$\sup_k \mathsf{E}\left[\left.\Vert\xi_{k}\Vert^2\right|\mathcal{F}_{k}\right] 
< \infty$} for all $k$, and hence, if {$\sum_n a_n^2 < \infty$},
\begin{align*}
\sum_{n=0}^{\infty}\mathsf{E}\left[\Vert{M_{n+1}-M_n}\Vert^2\right]
&= \sum_{n=0}^{\infty} a_n^2\mathsf{E}\left[\Vert\xi_{n}\Vert^2\right]
\leqslant \sum_{n=0}^{\infty} a_n^2 \sup_n \mathsf{E}\left[\Vert\xi_{n}\Vert^2\right]
< \infty \text{ a.s.}
\end{align*}
The claim follows from martingale convergence theorem~\cite[page 111]{Williams_1991_book_Cambridge}.
\qed

\begin{proposition}
\label{grad_SF1_convergence}
For a given {$q<\big(1+\frac{2}{N}\big)$}, {$q\neq1$}, and for all {$\theta\in C$}, the error term
\begin{displaymath}
\left\Vert \nabla_{\theta}(S_{q,\beta}^1 J)(\theta) - \nabla_{\theta}J(\theta)\right\Vert = o(\beta). 
\end{displaymath}
\end{proposition}
\proof
For small {$\beta>0$}, using Taylor series expansion of {$J(\theta+\beta\eta)$} around {$\theta\in C$},
\begin{displaymath}
J(\theta+\beta{\eta}) = J(\theta) + \beta{\eta}^T{\nabla_{\theta}J(\theta)} +
\frac{\beta^2}{2}{\eta}^T{\nabla_{\theta}^2 J(\theta)}\eta + o(\beta^2) .
\end{displaymath}
So we can write~\eqref{grad_SF1_formula} as 
\begin{align}
&\nabla_{\theta} (S_{q,\beta}^1 J)(\theta) \nonumber
= \frac{2}{(N+2-Nq)}\bigg(
\frac{J(\theta)}{\beta}\mathsf{E}_{G_q(\eta)}\left[\frac{\eta}{\rho{(\eta)}}\right] 
+ \mathsf{E}_{G_q(\eta)}\left[\frac{\eta\eta^T}{\rho{(\eta)}}\right]\nabla_{\theta}J(\theta) 
\\&\hspace{60mm}
+ \frac{\beta}{2}\mathsf{E}_{G_q(\eta)}\left[\left.\frac{\eta\eta^{T}\nabla_{\theta}^2J(\theta)\eta}{\rho{(\eta)}}\right|\theta\right]
+ o(\beta)\bigg)\;.
\label{GqSF1_expand}
\end{align}
We consider each term in~\eqref{GqSF1_expand}. The {$i^{th}$} component in the first term is
{$\mathsf{E}_{G_q(\eta)}\left[\frac{\eta^{(i)}}{\rho{(\eta)}}\right] =0$} by Proposition~\ref{moments}
for all {$i=1,\ldots,N$}.
Similarly, the {$i^{th}$} component in the third term can be written as
\begin{displaymath}
\frac{\beta}{2}\mathsf{E}_{G_q(\eta)}\left[\frac{\eta\eta^{T}\nabla_{\theta}^2J(\theta)\eta}{\rho{(\eta)}}\right]^{(i)} = 
\frac{\beta}{2}\sum\limits_{j=1}^{N}\sum\limits_{k=1}^{N}\left[\nabla_{\theta}^2J(\theta)\right]_{j,k}
\mathsf{E}_{G_q(\eta)}\left[\frac{\eta^{(i)}\eta^{(j)}\eta^{(k)}}{\rho{(\eta)}}\right].
\end{displaymath}
It can be observed that in all cases, each term in the summation is an odd function, and so from Proposition~\ref{moments},
we can show that the third term in~\eqref{GqSF1_expand} is zero. 
Using a similar argument, we claim that the off-diagonal terms in
{$\mathsf{E}_{G_q(\eta)}\left[\frac{\eta\eta^T}{\rho{(\eta)}}\right]$} are zero, while the diagonal terms are of the form
{$\mathsf{E}_{G_q(\eta)}\left[\frac{\left(\eta^{(i)}\right)^2}{\rho{(\eta)}}\right]$}, which exists for all 
{$q\in(-\infty,1)\bigcup(1,1+\frac{2}{N})$} as the conditions in Proposition~\ref{moments} are always satisfied on this interval.
Further, 
\begin{equation}
\mathsf{E}_{G_q(\eta)}\left[\frac{\left(\eta^{(i)}\right)^2}{\rho{(\eta)}}\right]
= \frac{(N+2-Nq)}{2}\;.
\label{grad_expand_term}
\end{equation}
The claim follows by substituting the above expression in~\eqref{GqSF1_expand}.
\qed

Now, we consider the following ODE for the slowest timescale recursion
\begin{equation}
\label{slowode}
\dot{\theta}(t) = \tilde{\mathcal{P}}_C \Big(\theta(t),\, -\nabla_{\theta}J(\theta(t))\Big),
\end{equation}
where {$\tilde{\mathcal{P}}_C$} is as defined in Lemma~\ref{kushner_thm_5.3.1}.
It can be
seen that the stationary points of \eqref{slowode} lie in the set 
$K = \big\{\theta\in C \big| \tilde{\mathcal{P}}_C \big(\theta,-\nabla_{\theta}J(\theta)\big) = 0 \big\}$.
%
%
We have the following key result which shows that iteration~\eqref{slowstep} tracks ODE~\eqref{slowode}.
%
%
\begin{theorem}
\label{thm_GqSF1}
Under Assumptions~\ref{ergodic} -- \ref{stepsize}, given {$\epsilon>0$} and 
{$q\in(-\infty,1)\bigcup(1,1+\frac{2}{N})$}, there exists {$\bar\beta >0$} such that
for all {$\beta\in(0,\bar\beta]$}, the sequence {$(\theta_n)_{n\geqslant0}$} obtained from G$q$-SF1 converges to the 
$\epsilon$-neighborhood of the stable attractor set of~\eqref{slowode}, defined as
$K^{\epsilon} = \{x : \Vert{x-x_0}\Vert < \epsilon, x_0\in K\}$
with probability 1 as {$n\to\infty$}.
\end{theorem}
\proof
It immediately follows from Lemmas~\ref{kushner_thm_5.3.1} and~\ref{xi_n_convergence} that 
the update in~\eqref{slowstep} converges to the stable fixed points of the ODE
\begin{equation}
\label{slowode_error}
\dot{\theta}(t) = \tilde{\mathcal{P}}_C \Big(\theta(t),\, -\nabla_{\theta}J(\theta(t)) + \Delta\big(\theta(t)\big)\Big)\;.
\end{equation}
Now starting from the same initial condition, the trajectory of~\eqref{slowode_error} 
converges to that of~\eqref{slowode} uniformly over compacts, as {$\Delta(\theta(t))\to0$}.
Since from Proposition~\ref{grad_SF1_convergence}, 
we have {$\left\Vert\Delta(\theta_n)\right\Vert = o(\beta)$} for all $n$,
the claim follows.
It may be noted that we can arrive at the same claim more technically using Hirsch's lemma~\cite{Hirsch_1989_jour_NeuNet}.
\qed
 
\noindent
{\bf Remark.} 
Under certain `richness' conditions on the noise,
one can show that, with probability 1, the above iterations 
can asymptotically avoid unstable stationary points and converge to the 
set of stable equilibria of~\eqref{slowode}~\cite[Chapter~4.3]{Bradiere_1998_jour_SIAMJCO,Borkar_2008_book_Cambridge}. 
However, in practice, due to the inherent randomness of the scheme, the recursions 
converge to the set of stable equilibria even without any additional noise conditions.

\subsection{Convergence of G$q$-SF2 Algorithm}

Since the proof of convergence here is along the lines of G$q$-SF1, 
we only provide a sketch of it. We just briefly describe the modifications
that are required in this case. Along the faster timescale, as {$n\to\infty$}, the updates
given by {$Z_{nL}$} track the function 
\begin{displaymath}
\left(\frac{\eta_n}{\beta(N+2-Nq)\rho{(\eta_n)}}\right)
\big(J(\theta_n+\beta\eta_n)-J(\theta_n-\beta\eta_n)\big). 
\end{displaymath}
So we can rewrite the slower timescale update for G$q$-SF2 algorithm, in a similar manner as~\eqref{slowstep},
where the noise term $\xi_n$ has two components, due to the two parallel simulations,
each being bounded (as in Lemma~\ref{xi_n_convergence}).
We have the following proposition for the error term
\begin{displaymath}
\Delta(\theta_n) = \nabla_{\theta_n}(S_{q,\beta}^2 J)(\theta_n) - \nabla_{\theta_n}J(\theta_n)\;.
\end{displaymath}

\begin{proposition}
\label{grad_SF2_convergence}
For a given {$q<\big(1+\frac{2}{N}\big)$}, {$q\neq1$}, and for all {$\theta\in C$},
\begin{displaymath}
\left\Vert \nabla_{\theta}(S_{q,\beta}^2 J)(\theta)-\nabla_{\theta}J(\theta)\right\Vert = o(\beta).
\end{displaymath}
\end{proposition}
One can prove the above claim by using Taylor's expasion of
$J(\theta+\beta\eta)-J(\theta-\beta\eta)$,
and arguments 
similar to Proposition~\ref{grad_SF1_convergence}.
%
%
Finally, we have the main convergence result for G$q$-SF2 algorithm.
\begin{theorem}
\label{thm_GqSF2}
Under Assumptions~\ref{ergodic} -- \ref{stepsize}, given {$\epsilon>0$} and 
{$q\in(-\infty,1)\bigcup(1,1+\frac{2}{N})$}, there exists {$\bar\beta >0$} such that
for all {$\beta\in(0,\bar\beta]$}, the sequence {$(\theta_n)_{n\geqslant0}$} obtained from G$q$-SF2 converges to
$\epsilon$-neighborhood of the stable attractor set of~\eqref{slowode}
almost surely as {$n\to\infty$}.
\end{theorem}

Theorems~\ref{thm_GqSF1} and~\ref{thm_GqSF2} provide conditions for convergence
of the algorithms to some {$\epsilon$}-neighborhood of some stable equilibria. 
This occurs due to the error term, which is $o(\beta)$. Hence,
to achieve $\epsilon$ arbitrarily small, one may decrease $\beta$ as the algorithm proceeds.
This is allowed as long as the sequences $\big(\frac{a_n}{\beta_n}\big)$ and $\big(\frac{b_n}{\beta_n}\big)$ satisfy Assumption~\ref{stepsize}.
The above results would still hold if $\bar\beta \geqslant \sup_n\beta_n = \beta_0$ (considering $\beta_n$ to be non-increasing sequence).
However, in practice, it is quite difficult to tune the rate of decrease of $\beta_n$ appropriately.
Moreover, these results do not give the precise value of {$\bar\beta$}. 
We also make a note on the analysis for Gaussian SF algorithms. Though the above results exclude
the case $q=1$, it is easy to verify that all the claims hold as $q\downarrow1$ 
as Proposition~\ref{moments} also holds in the limiting case.
Hence, the above convergence analysis also provides an alternative to the analysis
presented in~\cite{Bhatnagar_2007_jour_TOMACS} for Gaussian SF algorithms.

\section{Simulations using the Proposed Algorithms}
\label{sim_results}

\subsection{Numerical Setting}

\begin{figure}[h]
\centering
\includegraphics[height=20mm]{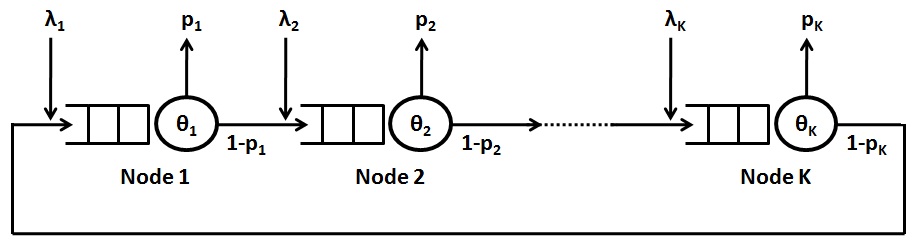}
\caption{Queuing Network.}
\label{fig_queue}
\end{figure}

We consider a multi-node network of {$M/G/1$} queues with feedback as shown in  Fig.~\ref{fig_queue}. 
There are $K$ nodes, which
are fed with independent Poisson external arrival processes with rates {$\lambda_1,\lambda_2,\ldots,\lambda_K$}, respectively.
After departing from the {$i^{th}$} node, a customer either leaves the system with probability {$p_i$} or enters the {$(i+1)^{th}$}
node with probability {$(1-p_i)$}. Once the service at the {$K^{th}$} node is completed, the customer may rejoin the {$1^{st}$} node
with probability {$(1-p_K)$}. The service time processes of each node, {$\{T_n^i(\theta^i)\}_{n\geqslant1}, i = 1,2,\ldots,K$} are defined as
\begin{equation}
T_n^i(\theta^i) = U_n^i \left({R_i+\Vert{\theta_n^i - \bar{\theta}^i}\Vert^2}\right),
\label{eq_serve_time}
\end{equation}
where {$R_i$}, {$i=1,2,\ldots,K$}, are constants and {$\{U_n^i : i=1,\ldots,K,n\geqslant0\}$} are independent samples drawn from the uniform distribution
on {$(0,1)$}. The service time of each node depends on the {$N_i$}-dimensional tunable parameter vector {$\theta^i$}, whose individual components
lie in a certain interval {$[\alpha_{min},\alpha_{max}]$}. 
{$\theta_n^i$} represents the {$n^{th}$} update of the parameter vector at the {$i^{th}$} node, and {$\bar{\theta}^i$}
represents the target parameter vector corresponding to the {$i^{th}$} node. 
The cost function is chosen to be the sum of the total waiting times of all the customers in the system. 
For the cost to be minimum, {$T_n^i(\theta^i)$} should
be minimum, and hence, we should have {$\theta_n^i=\bar{\theta}^i$}, {$i=1,\ldots,K$}. Let us denote 
{$\theta = (\theta^1, \theta^2, \ldots,\theta^K)^{T}$} and {$\bar\theta = (\bar\theta^1, \bar\theta^2, \ldots,\bar\theta^K)^{T}$}.
It is evident that {$\theta,\bar{\theta} \in\mathbb{R}^N$}, where {$N=\sum_{i=1}^{K} N_i$}. 
In order to compare the performance of the various
algorithms, we consider the performance measure to be the ratio of the Euclidean
distances of {$\theta_n$} and {$\theta_0$} from the vector {$\bar{\theta}$}, 
\textit{i.e.}, $\frac{\Vert\theta_n-\bar{\theta}\Vert}{\Vert\theta_0-\bar{\theta}\Vert}$.
The choice for such a performance measure is due to the fact that when the distance,
$\Vert\theta_n-\bar{\theta}\Vert$,
is low, the queuing network provides optimal performance. 
The ratio is considered to make the performance measure independent of the dimension 
of $\theta$. A low value of the ratio implies 
that the algorithm converges to a closer proximity of the global minimum. 
A similar measure was used by~\citeN{Spall_1992_jour_AutoControlTrans} to evaluate performance of SPSA methods.
All the results presented below are averaged over 100 independent runs. In addition to
the mean value of the estimated ratios (performance measure), we also indicate the 
coefficient of variation of the estimates (c.o.v.), which is basically the ratio of the standard
deviation and the mean of the estimates. This is used to indicate the consistency of the obtained results. 
The use of c.o.v. is more appropriate than standard deviation since the latter is often  
less for smaller mean.

\subsection{Experimental Results}

We begin with the simple case of a one-node network, where the service time
is controlled by a scalar parameter $\theta \in [0.1,0.6] \subset \mathbb{R}$.
The arrival rate at the node is {$\lambda_1 = 0.2$}, and the probability of leaving the 
system after service is {$p_1=0.4$}. We also fix the constant {$R_1=0.1$}, and 
the target parameter at {$\bar{\theta} = 0.3$}.
All simulations reported here were performed on an Intel Core {$i5$} machine with {$3.7GB$} memory space and Linux operating system. 
A total of 100 trials took about 10 seconds in this setting when we set $M=5000$ and $L=100$.

\begin{figure}[t]
\centering
\setlength{\tabcolsep}{1pt}
\begin{tabular}{|ccccc|}
\hline
Cauchy        & Cauchy       & Cauchy      & Gaussian     & $q=0$ \\
$\beta=0.01$ & $\beta=0.05$ & $\beta=0.15$ & $\beta=0.15$ & $\beta=0.15$ 
\\ \hline \hline
\includegraphics[width=20mm]{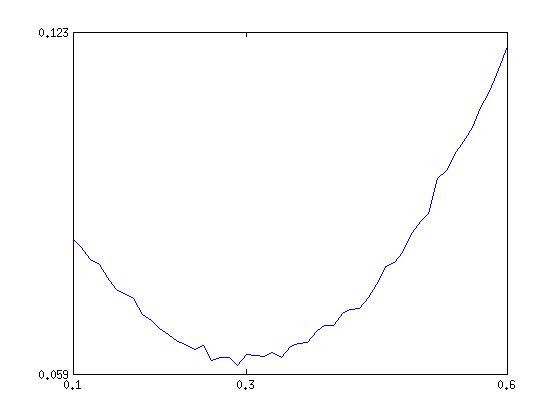} &
\includegraphics[width=20mm]{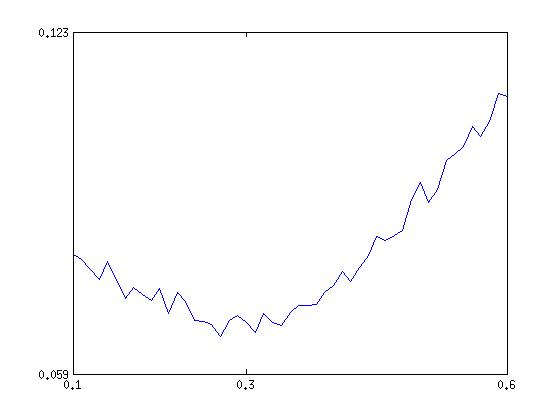} &
\includegraphics[width=20mm]{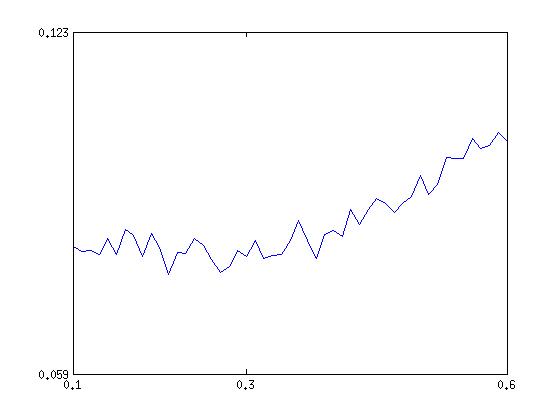} &
\includegraphics[width=20mm]{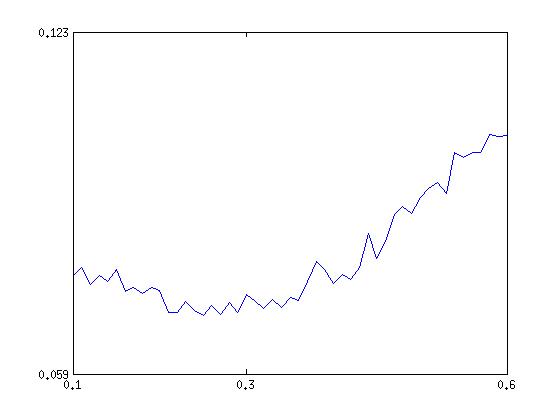} &
\includegraphics[width=20mm]{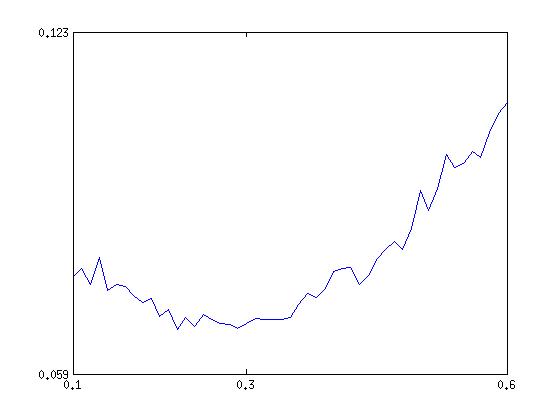} \\
\multicolumn{5}{|c|}{Estimated one-sided SF (y-axis) for varying $\theta$ (x-axis)}
\\ \hline
\includegraphics[width=20mm]{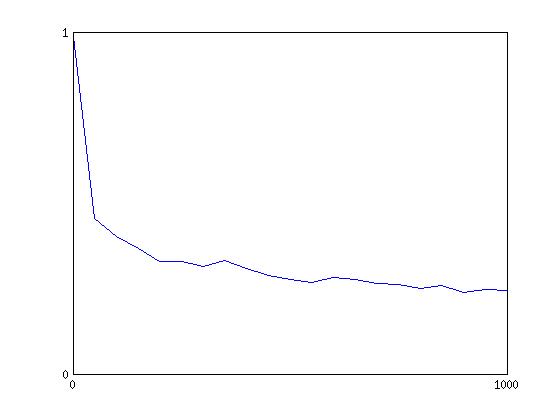} &
\includegraphics[width=20mm]{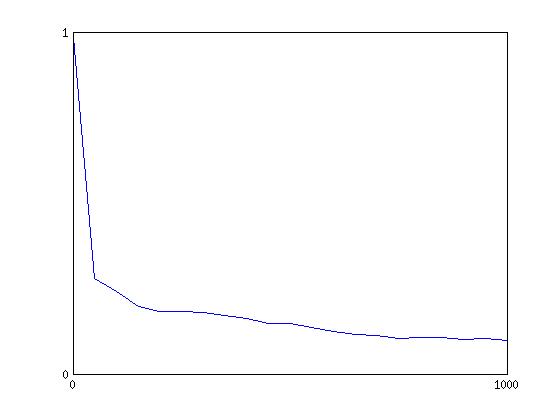} &
\includegraphics[width=20mm]{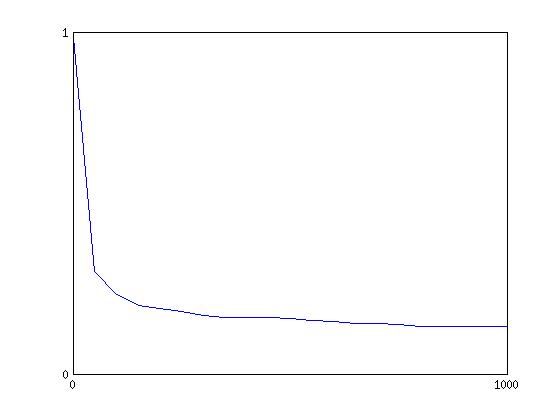} &
\includegraphics[width=20mm]{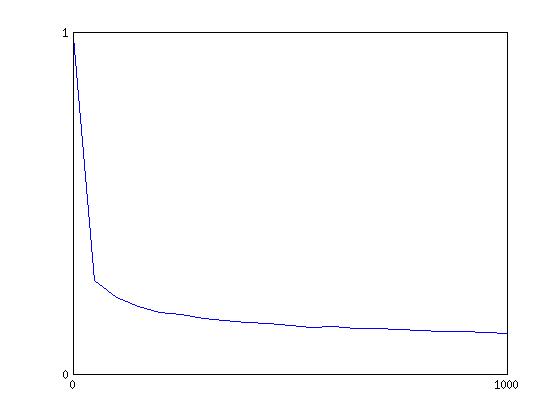} &
\includegraphics[width=20mm]{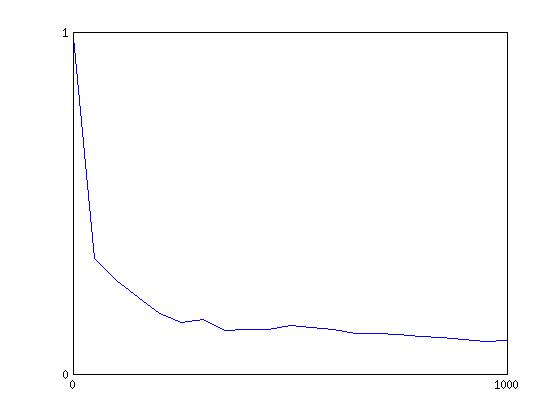} \\
\multicolumn{5}{|c|}{Convergence of G$q$-SF1: $\frac{\Vert\theta_n - \bar\theta\Vert}{\Vert\theta_0 - \bar\theta\Vert}$ (y-axis) vs. iterations (x-axis)}
\\ \hline \hline
\includegraphics[width=20mm]{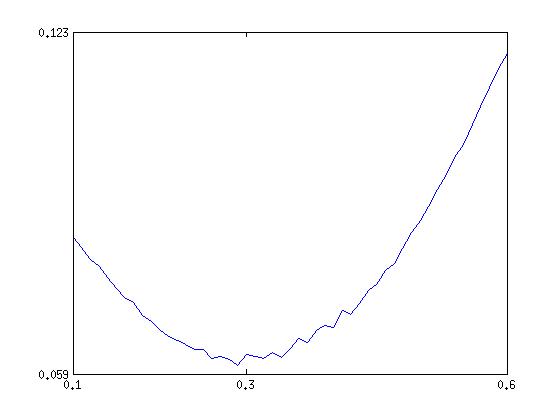} &
\includegraphics[width=20mm]{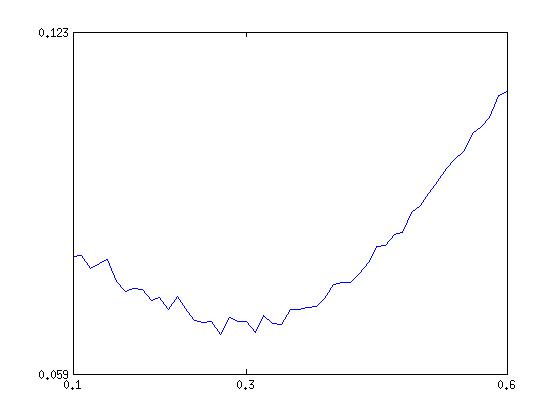} &
\includegraphics[width=20mm]{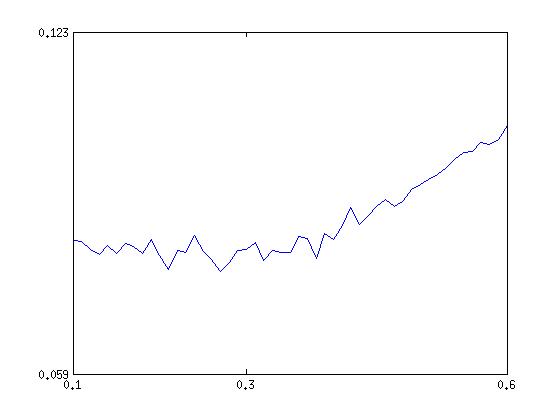} &
\includegraphics[width=20mm]{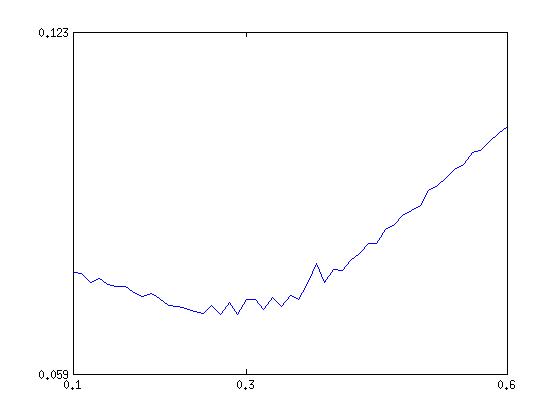} &
\includegraphics[width=20mm]{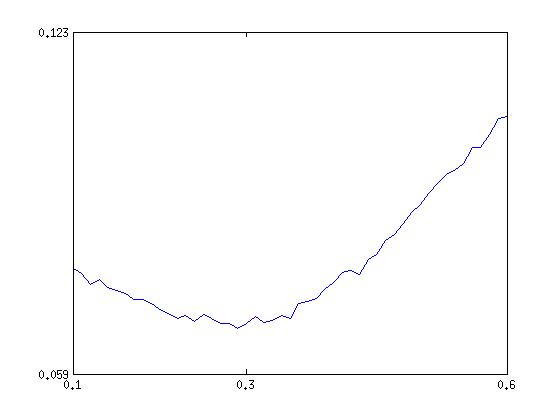} \\
\multicolumn{5}{|c|}{Estimated two-sided SF (y-axis) for varying $\theta$ (x-axis)}
\\ \hline
\includegraphics[width=20mm]{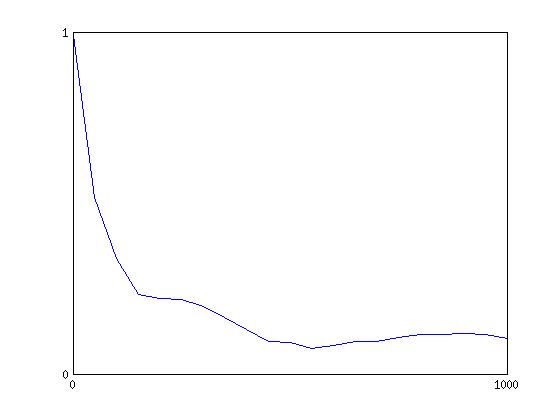} &
\includegraphics[width=20mm]{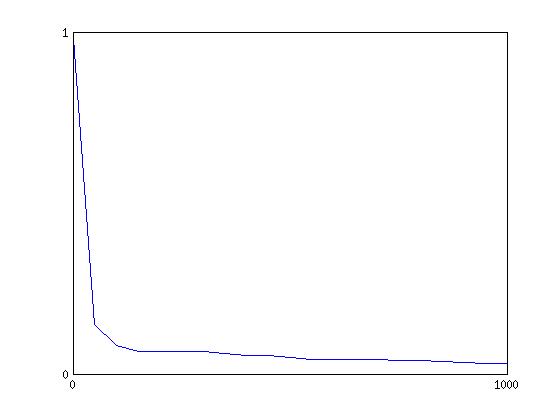} &
\includegraphics[width=20mm]{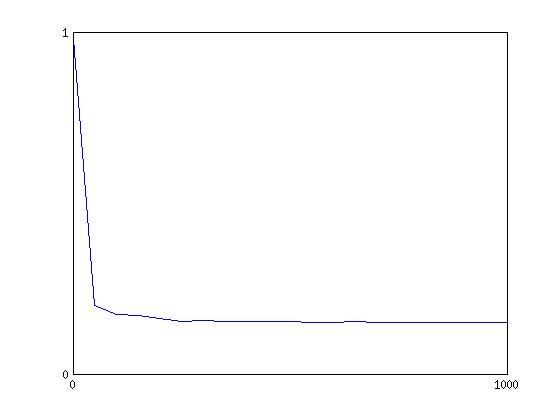} &
\includegraphics[width=20mm]{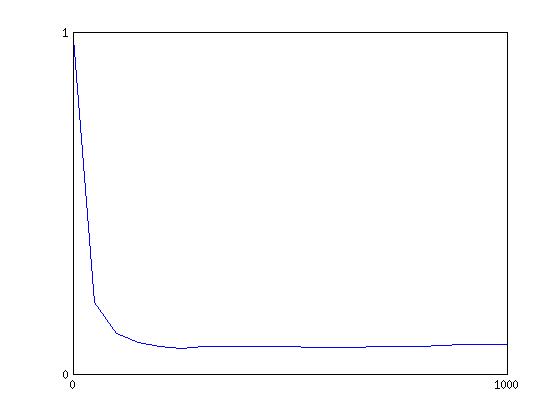} &
\includegraphics[width=20mm]{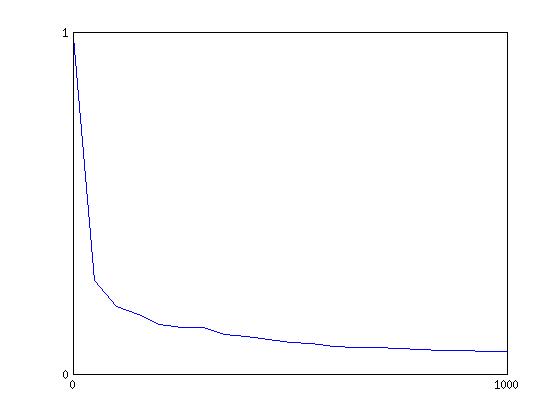} \\
\multicolumn{5}{|c|}{Convergence of G$q$-SF2: $\frac{\Vert\theta_n - \bar\theta\Vert}{\Vert\theta_0 - \bar\theta\Vert}$ (y-axis) vs. iterations (x-axis)}
\\ \hline
\end{tabular}
\caption{Nature of estimated SFs, and convergence behavior of G$q$-SF algorithms.}
\label{fig_sf_and_convergence}
\end{figure}

The effect of SFs is illustrated in
Fig.~\ref{fig_sf_and_convergence}. For this, we consider a few different kernels --
Cauchy ($q=2$), Gaussian ($q\to1$) and the kernel in random search method ($q=0$). 
The first and third rows show plots of the one and two-sided SFs, respectively, estimated 
for each $\theta\in[0.1,0.6]$ in steps of 0.01 using the relations,
\begin{align}
(S_{q,\beta}^1 J)(\theta) 
&= \mathsf{E}_{G_{q,\beta}(\eta)} \left[ \mathsf{E}_{\nu_{\theta-\eta}} [h(Y(\theta-\eta)) | \eta ]  \right]
\approx \frac{1}{TL}\sum_{n=1}^{T}\sum_{m=1}^{L} h(Y_m(\theta-\eta_n)) \;,
\\
(S_{q,\beta}^2 J)(\theta) 
&\approx \frac{1}{2TL}\sum_{n=1}^{T}\sum_{m=1}^{L} h(Y_m^1(\theta+\eta_n)) + h(Y_m^2(\theta-\eta_n)) \;.
\end{align}
We use $T=100$ independent trials with independent $q$-Gaussian variates $\eta_n$ and $L=10000$
simulations in each trial. The plots clearly indicate that fluctuations are less in the estimated two-sided SFs. They also demonstrate the effect of $q$ and $\beta$. The first
three columns consider Cauchy kernel with increasing $\beta$ values, and show how the 
SF becomes more flat for higher $\beta$. The last three columns consider same 
$\beta=0.15$, but three kernels with decreasing $q$ values ($q=2$, $q\to1$ and $q=0$), which indicate that the SFs becomes more flat as $q$ increases.

The second and fourth rows indicate the convergence of the G$q$-SF1 and G$q$-SF2
algorithms in $M=1000$ updates starting from $\theta_0=0.6$.
Here, we use only $L=100$ gradient updates on the 
faster timescale (as suggested in~\cite{Bhatnagar_2007_jour_TOMACS}). 
The plots, averaged over 100 runs, show that G$q$-SF2 exhibits better convergence 
behavior compared to G$q$-SF1, which can be directly attributed to the less jittery 
nature of the two-sided SFs. The effect of smoothing parameters, $q$ and $\beta$,
on the convergence behavior is quite obvious from the plots. 
Before discussing further about these parameters, we resolve the effect of the 
step-size sequence used in the two-timescale updates.

In the convergence plots of Fig.~\ref{fig_sf_and_convergence}, we considered the step-sizes 
to be $a_n=\frac{1}{n}$ and $b_n = \frac{1}{n^{0.75}}$. 
However, we may consider $b_n$ to be of the form $b_n=\frac{1}{n^{\gamma}}$,
where for any $\gamma\in(0.5,1)$, Assumption~\ref{stepsize} is satisfied.
For a smaller value of $\gamma$, the timescales will be well-separated and it appears 
that gradient updates converge much faster. This contrast reduces as $\gamma$ increases. 
In Table~\ref{tab_stepsize_1dim}, we study the effect of $\gamma$ for particular values of $q$ and $\beta$. 
Here, $q=1.0$ denotes Gaussian smoothing obtained in the limit of $q\to1$.
Since, Fig.~\ref{fig_sf_and_convergence} suggests that the algorithm almost converges 
to some $\theta$ in about 1000 iterations, we further perform our analysis 
using the value of 
$\frac{\Vert\theta_M-\bar\theta\Vert}{\Vert\theta_0-\bar\theta\Vert}$ where $M=5000$,
$L=100$ and $\theta_0=0.6$. 
The table shows this ratio averaged over 100 runs and the corresponding c.o.v. is shown in brackets.
For each $(q,\beta)$ pair, the minimum average ratio and the minimum c.o.v. are 
underlined. We can see that for $\gamma=0.75$ the minimum mean value
is achieved in most cases (9 of 20), and the c.o.v. is better when compared to 
other values of $\gamma$. Hence, for subsequent experiments, we use the step-size 
$b_n =\frac{1}{n^{0.75}}$.
One can also immediately observe that the ratio is much less in G$q$-SF2 as compared to G$q$-SF1. 
Though, the c.o.v. of both methods are comparable, there are few cases
where c.o.v. of G$q$-SF2 is considerably small. Among the 20 cases of different $(q,\beta)$ pairs,
G$q$-SF2 always achieves the minimum ratio value and the minimum c.o.v. is achieved in 15 cases. 
Hence, we observe that G$q$-SF2 exhibits better performance. Hence, the 
rest of the presented results are based on the G$q$-SF2 algorithm. Similar experiments 
were performed with G$q$-SF1 and the observed trends were similar to the ones for
G$q$-SF2 discussed below.

\begin{table}[b]
\centering
\scriptsize
\setlength{\tabcolsep}{3pt}
\setlength{\extrarowheight}{1pt}
\begin{tabular}{|c|c|c|c|c|c|c|c|c|c|}
\hline
&	&	\multicolumn{2}{c|}{$\gamma=0.55$}	&	\multicolumn{2}{c|}{$\gamma=0.65$}	&	\multicolumn{2}{c|}{$\gamma=0.75$}	&	\multicolumn{2}{c|}{$\gamma=0.85$}	\\
\cline{3-10}
$\beta$ 
&	$q$	&	G$q$-SF1	&	G$q$-SF2	&	G$q$-SF1	&	G$q$-SF2	&	G$q$-SF1	&	G$q$-SF2	&	G$q$-SF1	&	G$q$-SF2	\\
\hline
\multirow{5}{*}{\rotatebox{90}{$\beta=0.01$}}
&	-5.0	&	0.449 (0.67)	&	0.498 (\underline{0.58})	&	0.422 (0.65)	&	0.337 (0.70)	&	0.441 (0.60)	&	\underline{0.290} (0.71)	&	0.484 (0.60)	&	0.299 (0.77)	\\
&	0.0	&	0.356 (0.69)	&	0.204 (0.66)	&	0.323 (0.74)	&	0.127 (1.04)	&	0.333 (0.68)	&	\underline{0.174} (\underline{0.63})	&	0.402 (\underline{0.63})	&	0.191 (0.78)	\\
&	1.0	&	0.250 (0.76)	&	0.129 (0.73)	&	0.307 (0.74)	&	0.113 (0.79)	&	0.286 (0.71)	&	\underline{0.104} (\underline{0.56})	&	0.292 (0.70)	&	0.129 (0.66)	\\
&	2.0	&	0.248 (0.72)	&	0.071 (0.70)	&	0.238 (0.84)	&	0.088 (\underline{0.64})	&	0.207 (0.83)	&	\underline{0.068} (0.98)	&	0.235 (0.70)	&	0.072 (0.87)	\\
&	2.75	&	0.254 (0.79)	&	0.114 (0.81)	&	0.277 (0.76)	&	0.138 (0.76)	&	0.245 (0.81)	&	\underline{0.105} (0.68)	&	0.266 (0.78)	&	0.163 (\underline{0.58})	\\
\cline{1-2}
\multirow{5}{*}{\rotatebox{90}{$\beta=0.05$}}
&	-5.0	&	0.435 (0.63)	&	0.287 (0.74)	&	0.445 (0.53)	&	\underline{0.263} (0.69)	&	0.438 (\underline{0.55})	&	0.290 (0.76)	&	0.403 (0.69)	&	0.316 (0.90)	\\
&	0.0	&	0.250 (0.69)	&	0.139 (\underline{0.65})	&	0.261 (0.77)	&	0.120 (0.71)	&	0.274 (0.74)	&	\underline{0.103} (\underline{0.65})	&	0.252 (0.71)	&	0.106 (0.78)	\\
&	1.0	&	0.165 (0.80)	&	0.049 (0.70)	&	0.201 (0.77)	&	0.061 (\underline{0.61})	&	0.187 (0.68)	&	0.045 (0.72)	&	0.172 (0.86)	&	\underline{0.040} (0.68)	\\
&	2.0	&	0.155 (0.77)	&	\underline{0.037} (0.95)	&	0.138 (0.84)	&	0.044 (0.70)	&	0.137 (0.84)	&	0.045 (0.78)	&	0.150 (0.91)	&	0.051 (\underline{0.67})	\\
&	2.75	&	0.241 (0.78)	&	0.100 (0.88)	&	0.201 (0.84)	&	0.086 (0.76)	&	0.219 (\underline{0.71})	&	0.078 (0.75)	&	0.209 (0.80)	&	\underline{0.069} (0.77)	\\
\cline{1-2}
\multirow{5}{*}{\rotatebox{90}{$\beta=0.10$}}
&	-5.0	&	0.270 (0.86)	&	0.073 (0.91)	&	0.301 (0.71)	&	0.100 (1.29)	&	0.296 (0.76)	&	0.093 (0.89)	&	0.323 (\underline{0.67})	&	\underline{0.068} (0.73)	\\
&	0.0	&	0.080 (0.77)	&	0.023 (0.90)	&	0.079 (0.82)	&	\underline{0.021} (\underline{0.51})	&	0.100 (0.78)	&	0.023 (0.88)	&	0.082 (0.93)	&	0.022 (0.56)	\\
&	1.0	&	0.039 (0.81)	&	0.012 (0.74)	&	0.043 (0.67)	&	0.012 (0.67)	&	0.045 (\underline{0.66})	&	0.019 (0.68)	&	0.046 (0.70)	&	\underline{0.009} (0.72)	\\
&	2.0	&	0.053 (0.75)	&	0.034 (\underline{0.35})	&	0.048 (0.66)	&	\underline{0.031} (0.38)	&	0.044 (0.75)	&	0.033 (0.40)	&	0.049 (0.67)	&	0.035 (0.46)	\\
&	2.75	&	0.120 (0.72)	&	0.085 (0.83)	&	0.124 (\underline{0.71})	&	0.079 (0.77)	&	0.124 (0.78)	&	\underline{0.064} (0.74)	&	0.122 (0.81)	&	0.092 (0.81)	\\
\cline{1-2}
\multirow{5}{*}{\rotatebox{90}{$\beta=0.25$}}
&	-5.0	&	0.277 (\underline{0.79})	&	0.087 (1.64)	&	0.252 (0.84)	&	0.053 (1.15)	&	0.225 (0.90)	&	0.112 (1.96)	&	0.252 (0.83)	&	\underline{0.042} (0.81)	\\
&	0.0	&	0.052 (0.74)	&	0.020 (1.12)	&	0.056 (0.76)	&	\underline{0.011} (0.92)	&	0.057 (0.80)	&	0.013 (\underline{0.70})	&	0.055 (0.93)	&	0.019 (0.77)	\\
&	1.0	&	0.036 (0.72)	&	0.018 (\underline{0.40})	&	0.035 (0.72)	&	\underline{0.017} (0.50)	&	0.034 (0.70)	&	\underline{0.017} (0.53)	&	0.037 (0.64)	&	0.019 (0.45)	\\
&	2.0	&	0.081 (0.53)	&	0.084 (0.28)	&	0.074 (0.60)	&	0.073 (0.21)	&	0.081 (0.60)	&	\underline{0.070} (0.23)	&	0.078 (0.53)	&	0.073 (\underline{0.17})	\\
&	2.75	&	0.154 (0.74)	&	0.138 (0.79)	&	0.148 (0.77)	&	0.130 (0.74)	&	0.159 (0.82)	&	0.111 (0.82)	&	0.158 (0.78)	&	\underline{0.101} (\underline{0.49})	\\
\hline
\end{tabular}
\caption{Performance of algorithms for various $q$ and $\beta$ with 
step-sizes $a_n=\frac{1}{n}$, $b_n=\frac{1}{n^{\gamma}}$.}
\label{tab_stepsize_1dim}
\end{table}

We arrive at the main segment of our discussion -- the effect of choosing the 
appropriate smoothing kernel. Table~\ref{tab_kernel} provides a list of kernels and 
how these kernels can be retrieved from the $q$-Gaussian distribution. 
So the choice of kernel is essentially same as the choice of the parameter $q$.
Table~\ref{tab_GqSF2_1dim} provides the mean and c.o.v. of the ratio after
$M=5000$ iterations for a wide range of $q$ and $\beta$ values. Here, we use
the previously described one-dimensional setting with constants mentioned before.
We note that while any $\beta>0$ is applicable, $q$ is restricted to $q<(1+\frac{2}{N})=3$.
The case of G-SF2~\cite{Bhatnagar_2007_jour_TOMACS} is denoted by $q=1.0$,
while Cauchy smoothing is for $q=2.0$, random search for $q=0.0$ and very low values
of $q$, such as $q=-10$, yield performance quite similar to uniform smoothing.
The results presented in Table~\ref{tab_GqSF2_1dim} show that 
the best performance in the mean sense is observed for $q=0.5$ and $\beta=0.075$.
One can also observe the trend that as $\beta$ increases, smaller values of $q$ provide better convergence.
At the same time, results for larger $q$ and $\beta$ values seem to be more consistent as the c.o.v. is smaller
in such cases. We see that, in this particular setting, the performance of the algorithm
is quite good for $\beta\in[0.025,0.1]$ and $q\in[-0.5,2]$.
The c.o.v. in this range is also not very high, but one may prefer $\beta$ to be close to 0.1 to achieve
consistent performance, and hence, choosing $q<1$ may prove
more efficient in such a scenario. For G$q$-SF1 algorithm, similar nature of results were observed (not presented here).

\begin{table}[b]
\centering
\scriptsize
\setlength{\tabcolsep}{3pt}
\setlength{\extrarowheight}{1pt}
\begin{tabular}{|c||c|c|c|c|c|c|c|c|}
\hline
\backslashbox{$q$}{$\beta$}	&	0.0075	&	0.01	&	0.025	&	0.05	&	0.075	&	0.1	&	0.15	&	0.2	\\
\hline													
\hline													
-10.0	&	0.382 (0.57)	&	0.283 (0.72)	&	0.148 (0.83)	&	0.107 (0.97)	&	0.107 (0.80)	&	0.084 (0.63)	&	0.107 (1.39)	&	0.114 (0.97)	\\
-5.0	&	0.314 (0.63)	&	0.318 (0.88)	&	0.171 (0.75)	&	0.057 (1.17)	&	0.076 (1.03)	&	0.043 (1.36)	&	0.042 (0.58)	&	0.121 (0.89)	\\
-2.5	&	0.240 (1.02)	&	0.209 (0.78)	&	0.098 (0.96)	&	0.062 (0.70)	&	0.049 (1.09)	&	0.056 (0.92)	&	0.067 (1.60)	&	0.087 (0.96)	\\
-1.0	&	0.158 (0.75)	&	0.175 (0.67)	&	0.056 (0.75)	&	0.071 (1.48)	&	\underline{0.018} (0.89)	&	0.044 (1.80)	&	0.038 (1.51)	&	0.140 (0.35)	\\
-0.5	&	0.111 (0.76)	&	0.148 (0.91)	&	0.055 (0.69)	&	\underline{0.025} (0.67)	&	\underline{0.020} (0.72)	&	\underline{0.018} (1.13)	&	0.030 (0.59)	&	0.143 (0.41)	\\
-0.25	&	0.128 (0.91)	&	0.128 (0.65)	&	0.044 (0.79)	&	0.028 (0.60)	&	\underline{0.024} (0.73)	&	\underline{0.017} (0.64)	&	0.027 (0.74)	&	0.162 (0.26)	\\
0.0	&	0.123 (0.70)	&	0.121 (0.87)	&	0.044 (0.81)	&	\underline{0.025} (0.59)	&	\underline{0.025} (0.69)	&	\underline{0.010} (0.90)	&	0.043 (0.38)	&	0.164 (\underline{0.20})	\\
0.25	&	0.099 (0.60)	&	0.115 (1.13)	&	\underline{0.022} (0.78)	&	\underline{0.019} (0.56)	&	\underline{0.012} (0.78)	&	\underline{0.011} (0.71)	&	0.053 (0.42)	&	0.179 (\underline{0.24})	\\
0.5	&	0.078 (0.70)	&	0.070 (0.87)	&	0.037 (0.61)	&	\underline{0.019} (0.70)	&	\underline{0.006} (0.80)	&	\underline{0.009} (0.67)	&	0.063 (\underline{0.20})	&	0.203 (\underline{0.22})	\\
0.75	&	0.079 (0.78)	&	0.054 (0.61)	&	0.033 (0.62)	&	\underline{0.015} (0.81)	&	\underline{0.009} (0.83)	&	\underline{0.009} (1.23)	&	0.079 (\underline{0.19})	&	0.221 (\underline{0.21})	\\
1.0	&	0.081 (0.59)	&	0.057 (0.99)	&	\underline{0.020} (0.56)	&	\underline{0.009} (0.46)	&	\underline{0.009} (0.81)	&	\underline{0.018} (0.55)	&	0.097 (\underline{0.14})	&	0.229 (\underline{0.16})	\\
1.25	&	0.062 (0.75)	&	0.047 (0.79)	&	\underline{0.018} (0.61)	&	\underline{0.013} (0.82)	&	\underline{0.014} (0.86)	&	0.034 (0.40)	&	0.109 (\underline{0.23})	&	0.202 (0.39)	\\
1.5	&	0.060 (0.96)	&	0.046 (0.61)	&	\underline{0.014} (0.63)	&	\underline{0.013} (0.87)	&	0.025 (\underline{0.23})	&	0.047 (\underline{0.23})	&	0.135 (\underline{0.27})	&	0.191 (0.39)	\\
1.75	&	0.059 (0.49)	&	0.043 (0.82)	&	\underline{0.023} (0.76)	&	\underline{0.017} (0.60)	&	0.041 (0.26)	&	0.064 (\underline{0.20})	&	0.138 (\underline{0.17})	&	0.221 (0.30)	\\
2.0	&	0.064 (0.84)	&	0.042 (0.65)	&	\underline{0.017} (0.84)	&	0.027 (0.40)	&	0.055 (\underline{0.21})	&	0.073 (\underline{0.24})	&	0.136 (0.26)	&	0.165 (0.47)	\\
2.25	&	0.056 (0.73)	&	0.030 (0.71)	&	0.028 (0.56)	&	0.047 (0.31)	&	0.060 (0.36)	&	0.086 (0.44)	&	0.145 (0.49)	&	0.138 (0.65)	\\
2.5	&	0.055 (0.63)	&	0.049 (0.69)	&	0.035 (0.68)	&	0.047 (0.58)	&	0.050 (0.64)	&	0.107 (0.58)	&	0.092 (0.98)	&	0.130 (0.81)	\\
2.75	&	0.106 (0.71)	&	0.102 (0.74)	&	0.044 (0.90)	&	0.066 (0.60)	&	0.096 (0.76)	&	0.114 (0.66)	&	0.160 (0.74)	&	0.210 (0.58)	\\
\hline
\end{tabular}
\caption{Mean and c.o.v. of $\frac{\Vert\theta_M-\bar\theta\Vert}{\Vert\theta_0-\bar\theta\Vert}$ 
for G$q$-SF2 algorithm for different values of $q$ and $\beta$, considering the 1-dimensional setting.
Mean values $\leqslant 0.025$ and c.o.v. $\leqslant 0.25$ are underlined.}
\label{tab_GqSF2_1dim}
\end{table}

\begin{figure}[t]
\centering
\setlength{\tabcolsep}{1pt}
\begin{tabular}{|c|c|c|c|c|c|}
\hline
\backslashbox{$q$}{$\beta$} & $0.005$ & $0.01$ & $0.05$ & $0.15$ & $0.2$ \\ 
\hline 
-5.0 &
\includegraphics[width=20mm]{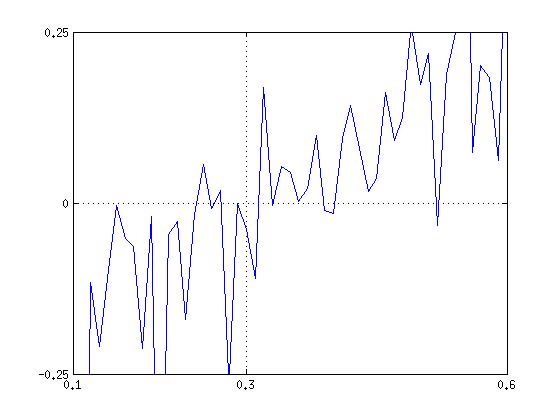} &
\includegraphics[width=20mm]{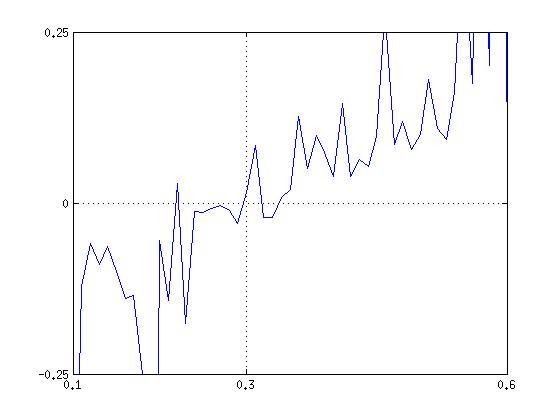} &
\includegraphics[width=20mm]{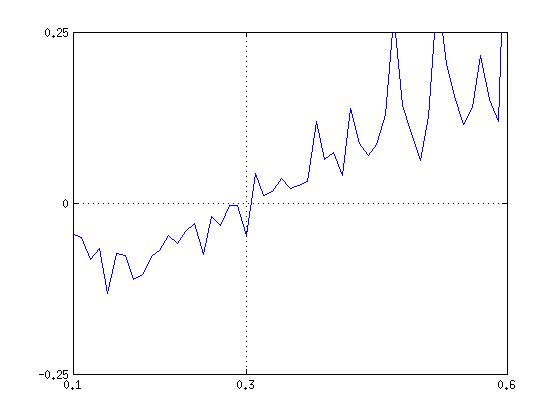} &
\includegraphics[width=20mm]{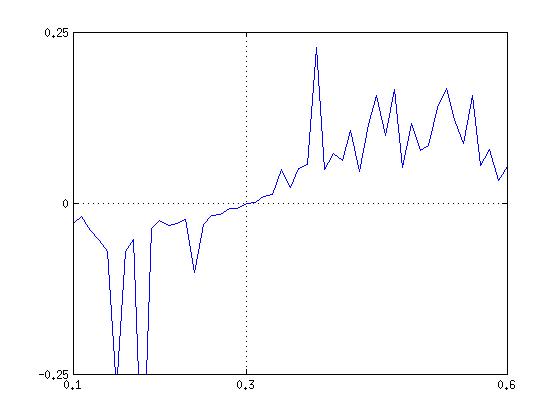} &
\includegraphics[width=20mm]{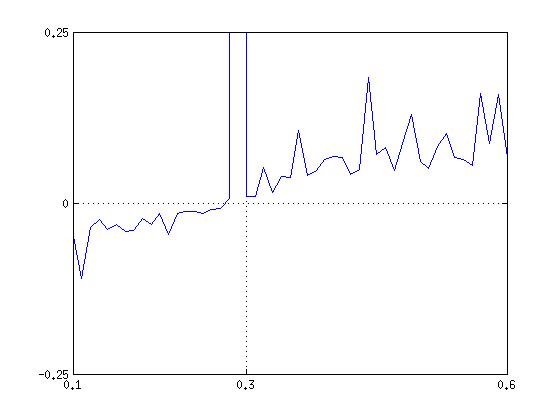} \\
\hline
0.0 &
\includegraphics[width=20mm]{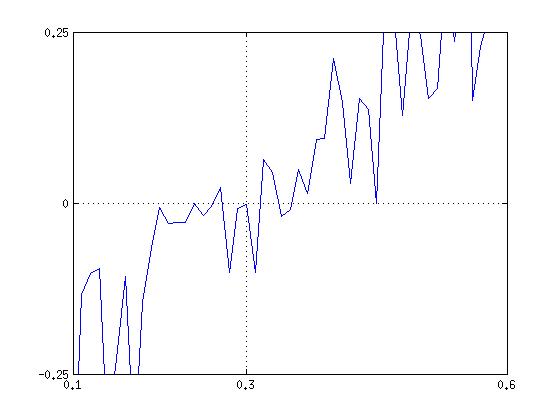} &
\includegraphics[width=20mm]{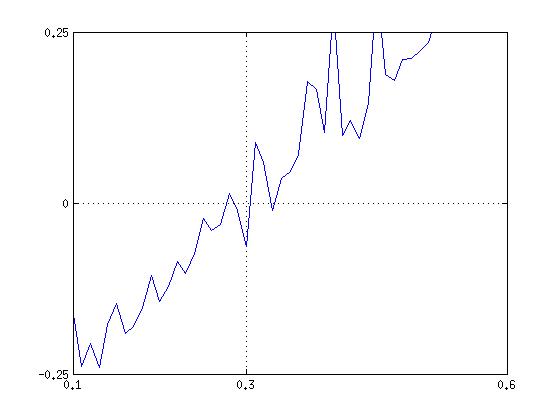} &
\includegraphics[width=20mm]{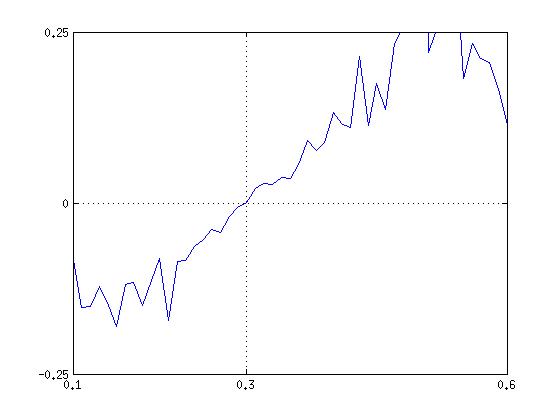} &
\includegraphics[width=20mm]{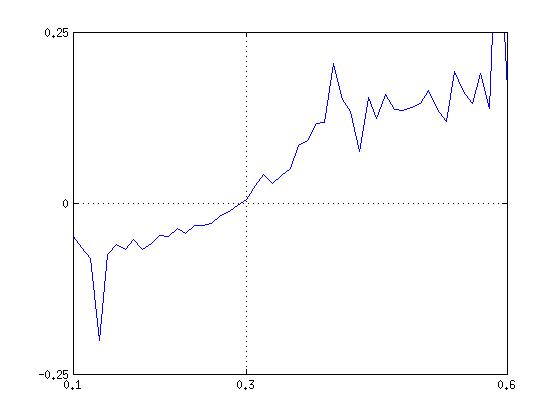} &
\includegraphics[width=20mm]{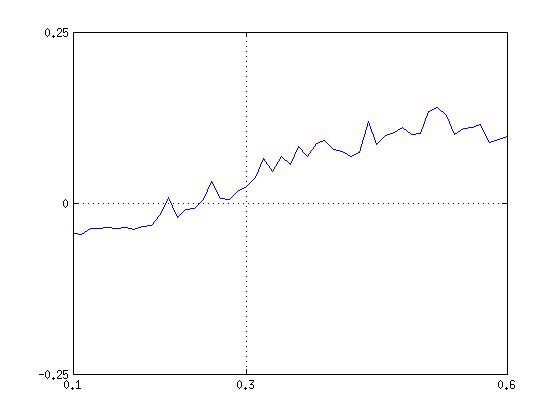} \\
\hline
1.0 &
\includegraphics[width=20mm]{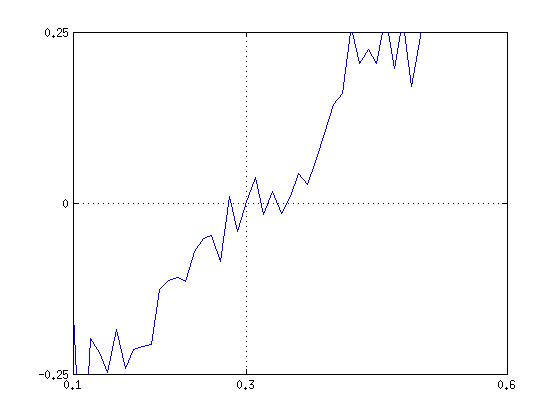} &
\includegraphics[width=20mm]{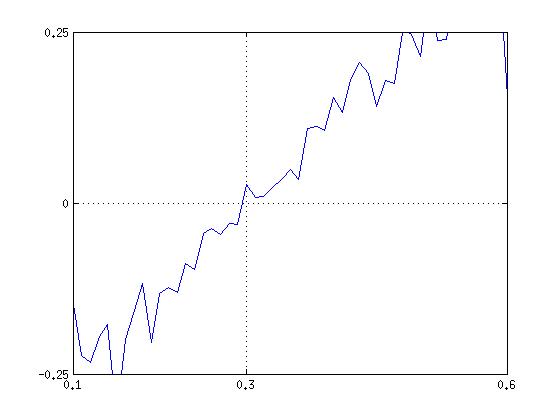} &
\includegraphics[width=20mm]{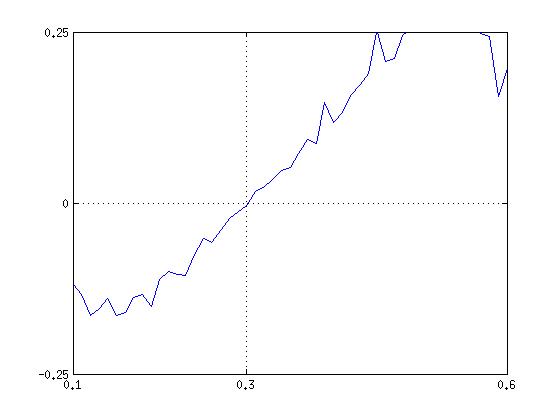} &
\includegraphics[width=20mm]{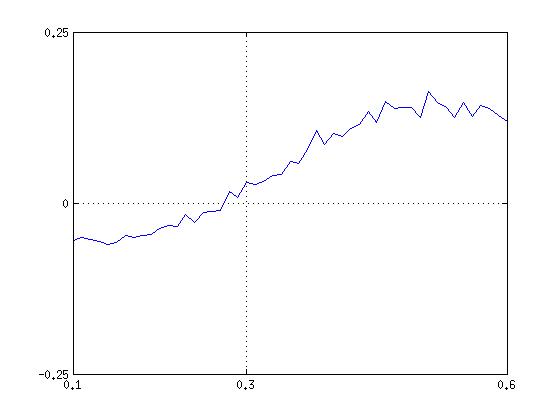} &
\includegraphics[width=20mm]{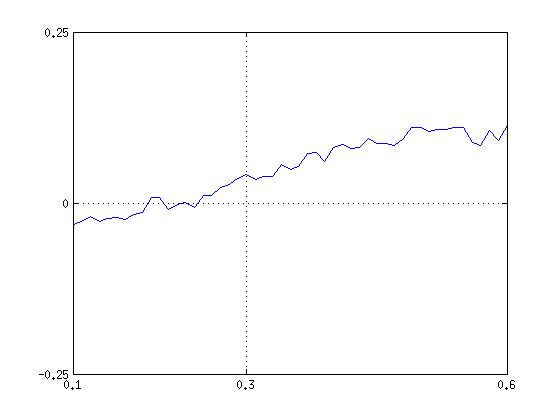} \\
\hline
2.0 &
\includegraphics[width=20mm]{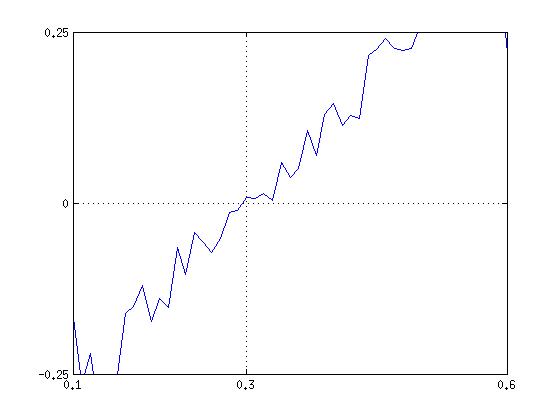} &
\includegraphics[width=20mm]{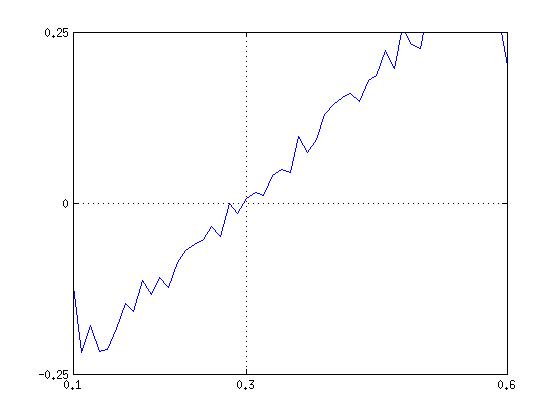} &
\includegraphics[width=20mm]{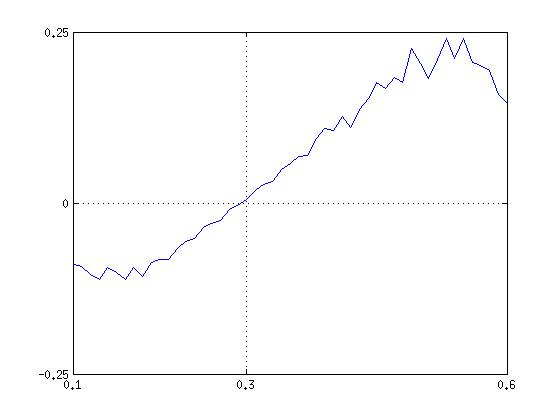} &
\includegraphics[width=20mm]{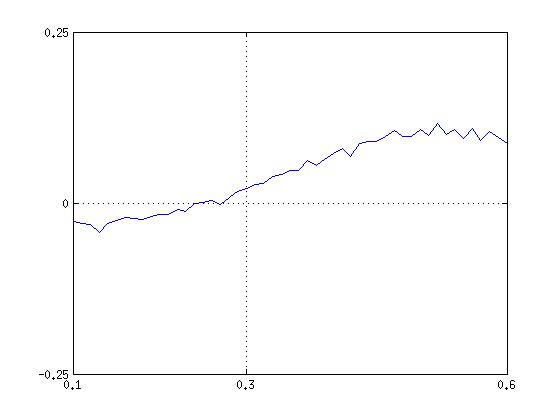} &
\includegraphics[width=20mm]{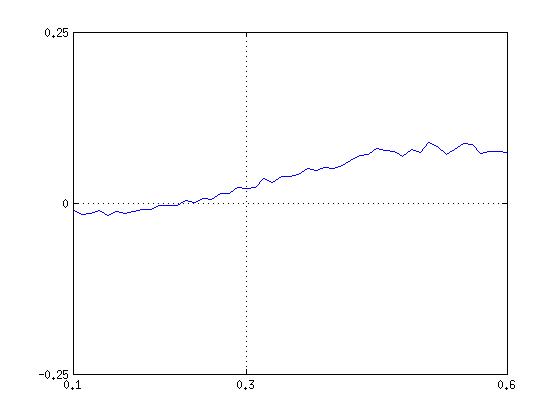} \\
\hline
2.75 &
\includegraphics[width=20mm]{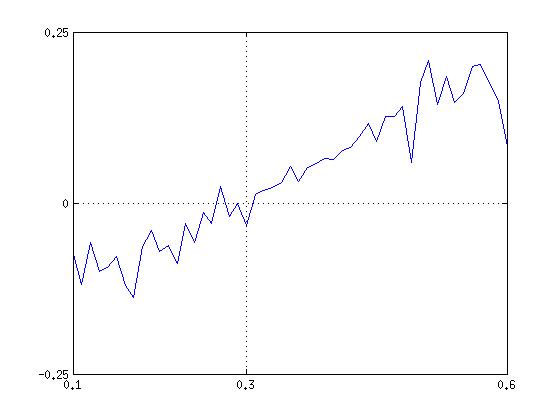} &
\includegraphics[width=20mm]{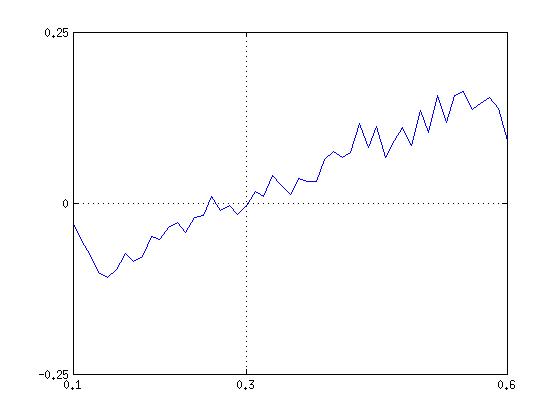} &
\includegraphics[width=20mm]{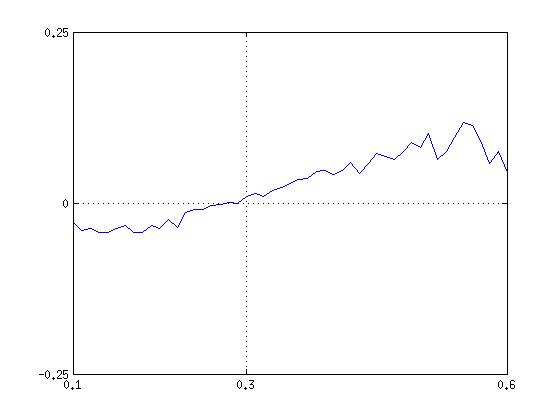} &
\includegraphics[width=20mm]{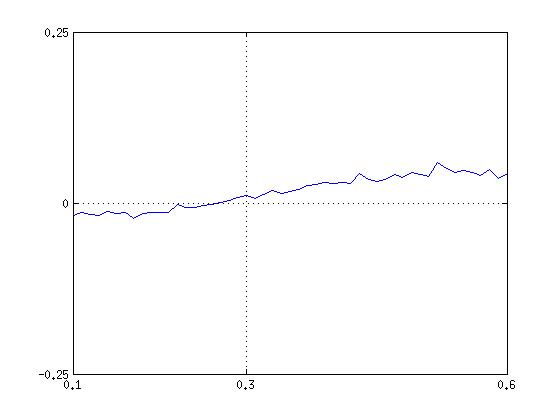} &
\includegraphics[width=20mm]{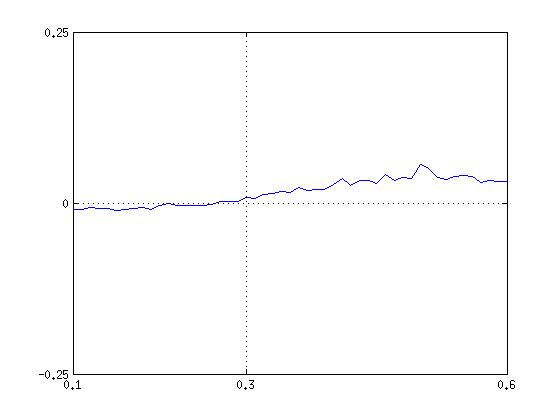} \\
\hline
\end{tabular}
\caption{Nature of estimated two-sided smoothed gradients (y-axis) for different parameter values $\theta$ (x-axis). 
Vertical line corresponds to $\theta=\bar\theta$, and horizontal one zero of gradient estimate, \textit{i.e.}, $Z=0$. All plots are drawn using the same scale for y-axis.}
\label{fig_grad}
\end{figure}

We now analyze the observations made in Table~\ref{tab_GqSF2_1dim}.
One may relate the nature of the SFs with the convergence behavior of the G$q$-SF2
algorithm for different values of $q$ and $\beta$.
But, it appears more justified to analyze the performance based on the basic idea
of stochastic optimization algorithms, \textit{i.e.}, finding the zeros of the estimated
gradient. The theoretical analysis in Section~\ref{GqSF_convergence}
guarantees convergence of the algorithm to a zero of the cost gradient.
Hence, the smoothing parameter should ideally be tuned such that a unique zero of
the gradient function is realized at the global optimum.
Though it is difficult to derive the optimal $(q,\beta)$ pair theoretically, we numerically
validate the fact that better performance is indeed achieved when the cost gradient
has a small number of zeros (may be only one) close to the global optimum,
which is $\bar\theta=0.3$ in the current setting.
Fig.~\ref{fig_grad} contains plots of the cost gradient estimated at different values of
$\theta$. The estimates are based on the relation in~\eqref{estimate2},
where we consider 100 independent trials, with the process in each trial being governed 
by a $q$-Gaussian perturbed parameter and running for $L=10000$ steps.
The plots in Fig.~\ref{fig_grad} indicate that for lower values of $q$ and $\beta$, the 
estimated gradient has a fluctuating nature and hence, has multiple zero crossings.
This results is relatively poor convergence and also higher c.o.v. since it may
converge to different local minima in different runs. As $\beta$ increases fluctuations
reduce, but this reduction (in fluctuations) is less prominent for lower values of $q$ ($q=-5$).
The middle column ($\beta=0.05$) exhibits cases ($q=0, 1$ and 2) with only one zero 
crossing at some $\theta$ very close to $\bar\theta$. This also corresponds to the 
better performance observed in Table~\ref{tab_GqSF2_1dim}. For larger values of $q$ and $\beta$,
the plots become smoother, but the zero crossing often occurs at some $\theta$ 
distant from $\bar\theta$. Further, the gradient becomes closer to zero for almost all
values of $\theta$, and hence, the steepest descent approach does not provide sufficient
exploration. Hence, the algorithm converges to some nearby optimum resulting in
less c.o.v., but poor performance (larger mean value).

We verify whether the above observations hold in a higher dimensional setting. 
For this, we consider a four node network with {$\lambda_i = 0.2$} and $p_i=0.2$ 
for all {$i=1,\ldots,4$}. The service process of each node is controlled by a $5$-dimensional 
parameter vector {$\theta^i\in[0.1,0.6]^5$}, a constant {$R_i=0.02$}
and a target value $\bar\theta^i = (0.3,0.3,0.3,0.3,0.3)^T$.
Thus, we have a $20$-dimensional parameter to be tuned.
We let {$M = 5000$}, {$L=100$} and
each component in the initial parameter vector, $\theta_0$, is at set at 0.6.
The chosen step-sizes are {$a_n = \frac{1}{n}$} and {$b_n = \frac{1}{n^{0.75}}$}.
For each {$(q,\beta)$} tuple, 100 independent runs were performed which took (entirely) about 80 seconds. 
We observe in Table~\ref{tab_GqSF2_20dim_waittime}
that again the best mean values are concentrated in the range 
$\beta\in[0.025,0.1]$ and $q\in[0,1.095]$, the upper limit coinciding with 20-dimensional Cauchy kernel. 
We also observe an overall improved c.o.v. in this case
compared to the 1-dimensional setting.
A similar set of experiments were performed where instead of the waiting
times of customers, we chose the single-stage cost functions to be the instantaneous
queue lengths. The results obtained in this experiment (not presented here)
were observed to be quite similar to the ones in Table~\ref{tab_GqSF2_20dim_waittime}.

To ensure that the above trends are not specific to the particular service time defined 
in~\eqref{eq_serve_time}, we modify the service time of each node 
$T_n^i(\theta^i)$ to be exponentially distributed with mean at 
$\Vert\theta_n^i-\bar\theta^i\Vert^2$, where a smaller mean ensures better performance. 
The other constants in the setting and the algorithm are same as before.
Table~\ref{tab_GqSF2_20dim_expserve} confirms that even in varying scenarios 
of the same queuing network setting, we can expect a reasonable performance
of the algorithm for $\beta\in[0.025,0.1]$ and $q\in\big[-0.5,1+\frac{2}{N+1}\big]$.
One may operate beyond this range, since the performance of the algorithm
decays gradually for both the continuous parameters $q$ and $\beta$, 
and a larger value of the latter may be preferred to ensure that c.o.v. is less.
However, if $\beta$ is too large, then the algorithms do not converge (Theorems~\ref{thm_GqSF1} and~\ref{thm_GqSF2}).
In simulations, it was observed that for large $\beta$ and $q$ close to $(1+\frac{2}{N})$, the simulation time increased drastically.

\begin{table}[t]
\centering
\scriptsize
\setlength{\tabcolsep}{3pt}
\setlength{\extrarowheight}{1pt}
\begin{tabular}{|c||c|c|c|c|c|c|c|c|}
\hline
\backslashbox{$q$}{$\beta$}	&	0.0075	&	0.01	&	0.025	&	0.05	&	0.075	&	0.1	&	0.15	&	0.2	\\
\hline													
\hline													
-10.0	&	0.545 (0.15)	&	0.499 (0.17)	&	0.372 (0.27)	&	0.291 (0.43)	&	0.253 (0.39)	&	0.248 (0.42)	&	0.333 (0.29)	&	0.369 (0.20)	\\
-5.0	&	0.522 (0.17)	&	0.497 (0.17)	&	0.340 (0.43)	&	0.238 (0.56)	&	0.206 (0.53)	&	0.207 (0.54)	&	0.313 (0.36)	&	0.348 (0.26)	\\
-2.5	&	0.467 (0.19)	&	0.451 (0.22)	&	0.222 (0.44)	&	0.141 (0.62)	&	0.132 (0.48)	&	0.150 (0.62)	&	0.234 (0.37)	&	0.318 (0.24)	\\
-1.0	&	0.460 (0.20)	&	0.378 (0.27)	&	0.150 (0.47)	&	0.077 (0.61)	&	0.079 (0.64)	&	0.094 (0.84)	&	0.185 (0.44)	&	0.285 (0.17)	\\
-0.5	&	0.381 (0.26)	&	0.305 (0.30)	&	0.103 (0.47)	&	0.049 (0.39)	&	0.048 (0.46)	&	0.061 (0.32)	&	0.153 (0.25)	&	0.272 (0.14)	\\
-0.25	&	0.357 (0.24)	&	0.260 (0.35)	&	0.079 (0.36)	&	0.041 (0.26)	&	0.039 (0.30)	&	0.049 (0.25)	&	0.141 (0.19)	&	0.253 (0.11)	\\
0.0	&	0.305 (0.30)	&	0.209 (0.38)	&	0.058 (0.28)	&	0.033 (0.22)	&	0.031 (0.33)	&	0.041 (0.24)	&	0.129 (0.15)	&	0.253 (\underline{0.10})	\\
0.25	&	0.249 (0.35)	&	0.159 (0.34)	&	0.046 (0.21)	&	0.026 (0.22)	&	\underline{0.025} (0.21)	&	0.034 (0.26)	&	0.118 (0.11)	&	0.249 (\underline{0.10})	\\
0.5	&	0.194 (0.34)	&	0.134 (0.37)	&	0.037 (0.25)	&	\underline{0.021} (0.21)	&	\underline{0.020} (0.17)	&	0.029 (0.17)	&	0.118 (0.11)	&	0.251 (\underline{0.08})	\\
0.75	&	0.151 (0.31)	&	0.104 (0.29)	&	0.031 (0.22)	&	\underline{0.018} (0.19)	&	\underline{0.017} (0.18)	&	0.027 (0.16)	&	0.112 (\underline{0.08})	&	0.241 (\underline{0.09})	\\
1.0	&	0.131 (0.28)	&	0.083 (0.28)	&	0.027 (0.24)	&	\underline{0.015} (0.16)	&	\underline{0.015} (0.16)	&	\underline{0.025} (0.13)	&	0.115 (\underline{0.08})	&	0.245 (\underline{0.08})	\\
1.02	&	0.125 (0.33)	&	0.087 (0.27)	&	\underline{0.025} (0.24)	&	\underline{0.015} (0.19)	&	\underline{0.015} (0.17)	&	0.028 (0.15)	&	0.117 (\underline{0.07})	&	0.249 (\underline{0.08})	\\
1.04	&	0.124 (0.30)	&	0.075 (0.25)	&	\underline{0.025} (0.17)	&	\underline{0.015} (0.17)	&	\underline{0.016} (0.19)	&	0.030 (0.16)	&	0.120 (\underline{0.08})	&	0.251 (\underline{0.07})	\\
1.06	&	0.123 (0.37)	&	0.079 (0.26)	&	\underline{0.025} (0.24)	&	\underline{0.015} (0.16)	&	\underline{0.017} (0.18)	&	0.035 (0.14)	&	0.129 (\underline{0.09})	&	0.256 (\underline{0.07})	\\
1.08	&	0.112 (0.30)	&	0.079 (0.33)	&	\underline{0.024} (0.19)	&	\underline{0.016} (0.14)	&	0.026 (0.14)	&	0.052 (\underline{0.10})	&	0.148 (\underline{0.09})	&	0.271 (\underline{0.09})	\\
1.095	&	0.130 (0.26)	&	0.094 (0.25)	&	\underline{0.023} (0.19)	&	0.059 (0.15)	&	0.084 (0.12)	&	0.117 (0.11)	&	0.200 (0.12)	&	0.282 (\underline{0.10})	\\
1.099	&	0.412 (0.16)	&	0.371 (0.14)	&	0.257 (0.15)	&	0.198 (0.18)	&	0.200 (0.16)	&	0.205 (0.17)	&	0.234 (0.14)	&	0.275 (0.12)	\\
\hline
\end{tabular}
\caption{Mean and c.o.v. of $\frac{\Vert\theta_M-\bar\theta\Vert}{\Vert\theta_0-\bar\theta\Vert}$ 
for G$q$-SF2 algorithm considering a 20-dimensional setting.
Mean values $\leqslant 0.025$ and c.o.v. $\leqslant 0.10$ are underlined.}
\label{tab_GqSF2_20dim_waittime}
\end{table}

\begin{table}[b]
\centering
\scriptsize
\setlength{\tabcolsep}{3pt}
\setlength{\extrarowheight}{1pt}
\begin{tabular}{|c||c|c|c|c|c|c|c|c|}
\hline
\backslashbox{$q$}{$\beta$}	&	0.0075	&	0.01	&	0.025	&	0.05	&	0.075	&	0.1	&	0.15	&	0.2	\\
\hline													
\hline													
-10	&	0.541 (0.17)	&	0.498 (0.20)	&	0.309 (0.49)	&	0.176 (0.66)	&	0.173 (0.68)	&	0.211 (0.59)	&	0.284 (0.31)	&	0.407 (0.27)	\\
-5	&	0.523 (0.19)	&	0.502 (0.23)	&	0.197 (0.64)	&	0.104 (0.91)	&	0.116 (0.67)	&	0.187 (0.61)	&	0.291 (0.40)	&	0.366 (0.24)	\\
-2.5	&	0.511 (0.20)	&	0.470 (0.25)	&	0.101 (0.82)	&	0.057 (0.76)	&	0.107 (1.02)	&	0.140 (0.97)	&	0.231 (0.45)	&	0.356 (0.26)	\\
-1	&	0.469 (0.27)	&	0.355 (0.46)	&	0.040 (0.84)	&	0.031 (0.84)	&	0.047 (0.78)	&	0.079 (0.82)	&	0.185 (0.47)	&	0.312 (0.18)	\\
-0.5	&	0.400 (0.37)	&	0.243 (0.58)	&	\underline{0.021} (0.63)	&	\underline{0.022} (0.51)	&	0.031 (0.47)	&	0.047 (0.39)	&	0.151 (0.20)	&	0.298 (0.14)	\\
-0.25&	0.349 (0.39)	&	0.177 (0.64)	&	\underline{0.017} (0.49)	&	\underline{0.015} (0.34)	&	\underline{0.025} (0.31)	&	0.046 (0.49)	&	0.139 (0.22)	&	0.290 (\underline{0.10})	\\
0	&	0.291 (0.47)	&	0.125 (0.63)	&	\underline{0.011} (0.41)	&	\underline{0.013} (0.35)	&	\underline{0.020} (0.21)	&	0.033 (0.22)	&	0.125 (0.14)	&	0.282 (\underline{0.10})	\\
0.25	&	0.199 (0.60)	&	0.090 (0.55)	&	\underline{0.009} (0.37)	&	\underline{0.010} (0.19)	&	\underline{0.017} (0.19)	&	0.030 (0.19)	&	0.118 (\underline{0.10})	&	0.275 (\underline{0.08})	\\
0.5	&	0.131 (0.57)	&	0.054 (0.51)	&	\underline{0.007} (0.33)	&	\underline{0.008} (0.16)	&	\underline{0.014} (0.18)	&	0.026 (0.16)	&	0.115 (\underline{0.10})	&	0.271 (\underline{0.07})	\\
0.75	&	0.089 (0.61)	&	0.038 (0.45)	&	\underline{0.005} (0.31)	&	\underline{0.008} (0.19)	&	\underline{0.012} (0.17)	&	\underline{0.025} (0.15)	&	0.112 (\underline{0.07})	&	0.266 (\underline{0.06})	\\
1.0	&	0.056 (0.48)	&	0.027 (0.55)	&	\underline{0.004} (0.28)	&	\underline{0.007} (0.16)	&	\underline{0.011} (0.14)	&	\underline{0.025} (0.16)	&	0.119 (\underline{0.07})	&	0.270 (\underline{0.05})	\\
1.02	&	0.059 (0.55)	&	0.026 (0.46)	&	\underline{0.004} (0.26)	&	\underline{0.007} (0.15)	&	\underline{0.012} (0.16)	&	0.027 (0.15)	&	0.121 (\underline{0.07})	&	0.273 (\underline{0.05})	\\
1.04	&	0.055 (0.60)	&	\underline{0.022} (0.45)	&	\underline{0.004} (0.23)	&	\underline{0.007} (0.15)	&	\underline{0.013} (0.15)	&	0.032 (0.12)	&	0.126 (\underline{0.06})	&	0.280 (\underline{0.06})	\\
1.06	&	0.053 (0.49)	&	\underline{0.025} (0.41)	&	\underline{0.004} (0.23)	&	\underline{0.008} (0.15)	&	\underline{0.016} (0.17)	&	0.039 (0.13)	&	0.140 (\underline{0.06})	&	0.293 (\underline{0.06})	\\
1.08	&	0.049 (0.49)	&	\underline{0.022} (0.56)	&	\underline{0.005} (0.21)	&	\underline{0.014} (0.18)	&	0.033 (0.13)	&	0.064 (0.11)	&	0.169 (\underline{0.07})	&	0.319 (\underline{0.05})	\\
1.095&	0.149 (0.32)	&	0.140 (0.34)	&	0.120 (0.23)	&	0.121 (0.19)	&	0.145 (0.16)	&	0.176 (0.13)	&	0.274 (\underline{0.08})	&	0.386 (\underline{0.06})	\\
1.099&	0.541 (0.12)	&	0.519 (0.14)	&	0.435 (0.15)	&	0.350 (0.14)	&	0.306 (0.14)	&	0.310 (0.15)	&	0.317 (0.13)	&	0.336 (0.11)	\\
\hline
\end{tabular}
\caption{Mean and c.o.v. of $\frac{\Vert\theta_M-\bar\theta\Vert}{\Vert\theta_0-\bar\theta\Vert}$ 
for G$q$-SF2 in a 20-dimensional setting with exponentially distributed service time.
Mean values $\leqslant 0.025$ and c.o.v. $\leqslant 0.10$ are marked.}
\label{tab_GqSF2_20dim_expserve}
\end{table}

\section{Conclusions}
\label{conclusion}
The origin and popularity of the $q$-Gaussian distribution is based on the notion that
it generalizes the Gaussian distribution.
This paper builds on this fact, and shows that it also generalizes most of the 
kernels studied in the context of smoothed functional algorithms.
Thus, tuning the parameter $q$ is equivalent to choosing the appropriate smoothing 
kernel. We have extended the smoothed functional approach for gradient estimation
to the $q$-Gaussian case. 
Based on this, we developed two-timescale gradient based search procedures that incorporate $q$-Gaussian smoothing. 
The convergence of the proposed methods to 
the stationary points of an associated ODE is proved.

The significance of $q$-Gaussian smoothing is demonstrated via numerical simulations
of a queuing network. The results show that tuning the parameter $q$ along with the 
smoothing parameter $\beta$ improves the performance of the algorithm in sense of 
both its convergence behavior and consistency. Appropriate ranges of $q$ and $\beta$
are provided, which do not vary much under various modifications of the setting considered. 
Developing rules for choosing $q$ and $\beta$, adaptively, would be an interesting future work.
We conclude by noting that the above idea can be extended
to derive Hessian estimators using $q$-Gaussian SFs, and developing Newton based algorithms along these lines.

\appendix
\section*{APPENDIX: Sampling algorithm for multivariate \lowercase{$q$}-Gaussian distribution}
\label{app_sampling}
\setcounter{section}{1}

The algorithms discussed in the paper require generation of $q$-Gaussian random vectors,
whose individual components are uncorrelated and identically distributed. 
For {$q\to1$} (Gaussian), uncorrelated implies independence of the components.
Hence, we can use standard algorithms to generate i.i.d. samples. This is typically not possible for 
$q$-Gaussians with {$q\neq1$}. \citeN{Thistleton_2007_jour_ITTrans} proposed an algorithm for generating 
one-dimensional $q$-Gaussian distributed random variables using generalized Box{-}M\"{u}ller transformation.
But, there exists no standard
algorithm for generating $N$-variate $q$-Gaussian random vectors. 
A method can be obtained by making use of the correspondence between $q$-Gaussian
and Students'-$t$ distributions for {$q>1$}. 
Further, a duality property of $q$-Gaussians can be used to relate the distributions for {$q\in\big(1,1+\frac{2}{N+2}\big)$}
and {$q\in(-\infty,1)$}. This observation, first made by~\citeN{Vignat_2006_jour_PhysicaA},
is stated below. 
Based on this, we present a method for sampling
$N$-variate $q$-Gaussian. 
We denote the $q$-Gaussian distribution with $q$-mean $\mu_q$ and
$q$-covariance $\Sigma_q$ as {$\mathcal{G}_q(\mu_q,\Sigma_q)$}.
\begin{lemma}
Let {$Y \sim \mathcal{G}_{q}(0,I_{N\times N})$} for some {$q\in\big(1,1+\frac{2}{N+2}\big)$}. Then the quantity 
\\$X =  \frac{\sqrt{\frac{2-q}{N+2-Nq}}Y}{\sqrt{1+\frac{q-1}{N+2-Nq}Y^{T}Y}} \sim \mathcal{G}_{q'}(0,I_{N\times N})$, where
{$q' = \left(1-\frac{q-1}{(N+4)-(N+2)q}\right)$}.
\end{lemma}

\begin{algorithm}[ht]
\caption{Sampling algorithm for multivariate $q$-Gaussian distribution}
\DontPrintSemicolon
\SetAlgoNoLine
\KwIn{
{$q\in\big(-\infty,1)\cup(1,1+\frac{2}{N}\big)$}, 
{$q$}-mean {$\mu_q\in\mathbb{R}^N$}, and
{$q$}-covariance matrix {$\Sigma_q\in\mathbb{R}^{N\times N}$}
}
Generate $N$-dimensional standard Gaussian vector {$Z \sim \mathcal{N}(0,I_{N\times N})$}.\;
Generate chi-squared random variate,
$a \sim \displaystyle\left\{\begin{array}{lcl}
		\chi^2\left(\frac{2(2-q)}{1-q}\right)
		&\text{for}  &-\infty<q<1,
		\vspace{2mm}\\
		\chi^2\left(\frac{N+2-Nq}{q-1}\right) 
		&\text{for} & 1<q<\left(1+\frac{2}{N}\right).
                \end{array}\right.$\;
Compute
 $Y = \displaystyle\left\{\begin{array}{rcl}
		\sqrt{\frac{N+2-Nq}{1-q}} \frac{Z}{\sqrt{a+Z^{T}Z}}
		&\text{for}  &-\infty<q<1,
		\vspace{2mm}\\
		\sqrt{\frac{N+2-Nq}{q-1}} \frac{Z}{\sqrt{a}}
		&\text{for} & 1<q<\left(1+\frac{2}{N}\right).
                \end{array}\right.$\;
\KwOut{{$X = \left(\mu_q + \Sigma_q^{1/2}Y\right)$}, which is a sample from {$\mathcal{G}_q(\mu_q,\Sigma_q)$}.}
\end{algorithm}

\bibliographystyle{ACM-Reference-Format-Journals}
\bibliography{qEntropy,StocApp}


\begin{thebibliography}{00}


\ifx \showCODEN    \undefined \def \showCODEN     #1{\unskip}     \fi
\ifx \showDOI      \undefined \def \showDOI       #1{{\tt DOI:}\penalty0{#1}\ }
  \fi
\ifx \showISBNx    \undefined \def \showISBNx     #1{\unskip}     \fi
\ifx \showISBNxiii \undefined \def \showISBNxiii  #1{\unskip}     \fi
\ifx \showISSN     \undefined \def \showISSN      #1{\unskip}     \fi
\ifx \showLCCN     \undefined \def \showLCCN      #1{\unskip}     \fi
\ifx \shownote     \undefined \def \shownote      #1{#1}          \fi
\ifx \showarticletitle \undefined \def \showarticletitle #1{#1}   \fi
\ifx \showURL      \undefined \def \showURL       #1{#1}          \fi

\bibitem[\protect\citeauthoryear{Bhatnagar}{Bhatnagar}{2007}]%
        {Bhatnagar_2007_jour_TOMACS}
{S. Bhatnagar}. 2007.
\newblock \showarticletitle{Adaptive {N}ewton-based multivariate smoothed
  functional algorithms for simulation optimization}.
\newblock {\em ACM Trans. Model. Comput. Simul.\/} {18}, 1 (2007), 27--62.
\newblock


\bibitem[\protect\citeauthoryear{Bhatnagar and Borkar}{Bhatnagar and
  Borkar}{1998}]%
        {Bhatnagar_1998_jour_ProbEnggInfoSc}
{S. Bhatnagar} {and} {V.~S. Borkar}. 1998.
\newblock \showarticletitle{Two timescale stochastic approximation scheme for
  simulation-based parametric optimization}.
\newblock {\em Prob. Engg. Info. Sci.\/}  {12} (1998), 519--531.
\newblock


\bibitem[\protect\citeauthoryear{Bhatnagar and Borkar}{Bhatnagar and
  Borkar}{2003}]%
        {Bhatnagar_2003_jour_Simulation}
{S. Bhatnagar} {and} {V.~S. Borkar}. 2003.
\newblock \showarticletitle{Multiscale chaotic {SPSA} and smoothed functional
  algorithms for simulation optimization}.
\newblock {\em Simulation\/} {79}, 9 (2003), 568--580.
\newblock


\bibitem[\protect\citeauthoryear{Borkar}{Borkar}{2008}]%
        {Borkar_2008_book_Cambridge}
{V.~S. Borkar}. 2008.
\newblock {\em Stochastic approximation: {A} dynamical systems viewpoint}.
\newblock Cambridge University Press.
\newblock


\bibitem[\protect\citeauthoryear{Brandiere}{Brandiere}{1998}]%
        {Bradiere_1998_jour_SIAMJCO}
{O. Brandiere}. 1998.
\newblock \showarticletitle{Some pathological traps for stochastic
  approximation}.
\newblock {\em SIAM J. Contr. and Optim.\/}  {36} (1998), 1293--1314.
\newblock


\bibitem[\protect\citeauthoryear{Fu}{Fu}{2006}]%
        {Fu_2006_conf_Simulation}
{M.~C. Fu}. 2006.
\newblock \showarticletitle{{G}radient estimation}.
\newblock In {\em Simulation}. Handbooks in Oper. Res. Man. Sci., Vol.~13.
  575--616.
\newblock


\bibitem[\protect\citeauthoryear{Ghoshdastidar, Dukkipati, and
  Bhatnagar}{Ghoshdastidar et~al\mbox{.}}{2012}]%
        {Ghoshdastidar_2012_conf_ISIT}
{D. Ghoshdastidar}, {A. Dukkipati}, {and} {S. Bhatnagar}. 2012.
\newblock \showarticletitle{{$q$}-{G}aussian based smoothed functional
  algorithms for stochastic optimization}. In {\em IEEE Int. Symp. Inf. Theory
  (ISIT)}.
\newblock


\bibitem[\protect\citeauthoryear{Gradshteyn and Ryzhik}{Gradshteyn and
  Ryzhik}{1994}]%
        {Gradshteyn_1994_book_Elesevier}
{I.~S. Gradshteyn} {and} {I.~M. Ryzhik}. 1994.
\newblock {\em Table of Integrals, Series and Products (5th ed.)}.
\newblock Elsevier.
\newblock


\bibitem[\protect\citeauthoryear{Hirsch}{Hirsch}{1989}]%
        {Hirsch_1989_jour_NeuNet}
{M.~W. Hirsch}. 1989.
\newblock \showarticletitle{Convergent activation dynamics is in continuous
  time networks}.
\newblock {\em Neural Networks\/}  {2} (1989), 331--349.
\newblock


\bibitem[\protect\citeauthoryear{Katkovnik and Kulchitsky}{Katkovnik and
  Kulchitsky}{1972}]%
        {Katkovnik_1972_jour_Automation}
{V.~Y.~A. Katkovnik} {and} {Y.~U. Kulchitsky}. 1972.
\newblock \showarticletitle{Convergence of a class of random search
  algorithms}.
\newblock {\em Auto. Remote Control\/}  {8} (1972), 1321--1326.
\newblock


\bibitem[\protect\citeauthoryear{Kiefer and Wolfowitz}{Kiefer and
  Wolfowitz}{1952}]%
        {Kiefer_1952_jour_AnnMathStat}
{E. Kiefer} {and} {J. Wolfowitz}. 1952.
\newblock \showarticletitle{Stochastic estimation of a maximum regression
  function}.
\newblock {\em Ann. Math. Statist.\/}  {23} (1952), 462--466.
\newblock


\bibitem[\protect\citeauthoryear{Kreimer and Rubinstein}{Kreimer and
  Rubinstein}{1988}]%
        {Kreimer_1988_jour_SIAMNumeAnal}
{J. Kreimer} {and} {R.~Y. Rubinstein}. 1988.
\newblock \showarticletitle{Smoothed functionals and constrained stochastic
  approximation}.
\newblock {\it SIAM J. Numer. Anal.} {25}, 2 (1988), 470--487.
\newblock


\bibitem[\protect\citeauthoryear{Kreimer and Rubinstein}{Kreimer and
  Rubinstein}{1992}]%
        {Kreimer_1992_jour_AnnOR}
{J. Kreimer} {and} {R.~Y. Rubinstein}. 1992.
\newblock \showarticletitle{Nondifferentiable optimization via smooth
  approximation: {G}eneral analytical approach}.
\newblock {\em Ann. Oper. Res.\/} {39}, 1 (1992), 97--119.
\newblock


\bibitem[\protect\citeauthoryear{Kushner and Clark}{Kushner and Clark}{1978}]%
        {Kushner_1978_book_Springer}
{H.~J. Kushner} {and} {D.~S. Clark}. 1978.
\newblock {\em Stochastic approximation methods for constrained and
  unconstrained systems}.
\newblock Springer-Verlag, New York.
\newblock


\bibitem[\protect\citeauthoryear{L'{E}cuyer and Glynn}{L'{E}cuyer and
  Glynn}{1994}]%
        {LEcuyer_1994_jour_ManSc}
{P. L'{E}cuyer} {and} {P.~W. Glynn}. 1994.
\newblock \showarticletitle{Stochastic optimization by simulation:
  {C}onvergence proofs for the {GI/G/1} queue in steady state}.
\newblock {\em Man. Sci.\/} {40}, 11 (1994), 1562--1578.
\newblock


\bibitem[\protect\citeauthoryear{Prato and Tsallis}{Prato and Tsallis}{1999}]%
        {Prato_1999_jour_PhyRevE}
{D. Prato} {and} {C. Tsallis}. 1999.
\newblock \showarticletitle{Nonextensive foundation of {L}\'{e}vy
  distributions}.
\newblock {\em Phy. Rev. E.\/}  {60} (1999), 2398.
\newblock


\bibitem[\protect\citeauthoryear{Robbins and Monro}{Robbins and Monro}{1951}]%
        {Robbins_1951_jour_AnnMathStat}
{H. Robbins} {and} {S. Monro}. 1951.
\newblock \showarticletitle{A stochastic approximation method}.
\newblock {\em Ann. Math. Statist.\/} {22}, 3 (1951), 400--7.
\newblock


\bibitem[\protect\citeauthoryear{Rubinstein}{Rubinstein}{1981}]%
        {Rubinstein_1981_book_Wiley}
{R.~Y. Rubinstein}. 1981.
\newblock {\em Simulation and {M}onte-{C}arlo method}.
\newblock John Wiley, New York.
\newblock


\bibitem[\protect\citeauthoryear{Ruppert}{Ruppert}{1985}]%
        {Ruppert_1985_jour_AnnStat}
{D. Ruppert}. 1985.
\newblock \showarticletitle{A {N}ewton-{R}aphson version of the multivariate
  {R}obbins-{M}onro procedure}.
\newblock {\em Ann. Statistics\/}  {13} (1985), 236--245.
\newblock


\bibitem[\protect\citeauthoryear{Sato}{Sato}{2010}]%
        {Sato_2001_jour_JourPhyConf}
{A.~H. Sato}. 2010.
\newblock \showarticletitle{\textit{q}-{G}aussian distributions and
  multiplicative stochastic processes for analysis of multiple financial time
  series}.
\newblock {\em J. Phys. Conf. Series\/} {201}, 012008 (2010).
\newblock


\bibitem[\protect\citeauthoryear{Schweitzer}{Schweitzer}{1968}]%
        {Schweitzer_1968_jour_AppProb}
{P.~J. Schweitzer}. 1968.
\newblock \showarticletitle{Perturbation theory and finite {M}arkov chains}.
\newblock {\em J. Appl. Prob.\/} (1968), 401--413.
\newblock


\bibitem[\protect\citeauthoryear{Spall}{Spall}{1992}]%
        {Spall_1992_jour_AutoControlTrans}
{J.~C. Spall}. 1992.
\newblock \showarticletitle{Multivariate stochastic approximation using a
  simultaneous perturbation gradient approximation}.
\newblock {\em IEEE Trans. Auto. Ctrl.\/} {37}, 3 (1992), 332--334.
\newblock


\bibitem[\protect\citeauthoryear{Spall}{Spall}{2000}]%
        {Spall_2000_jour_AutoCtrl}
{J.~C. Spall}. 2000.
\newblock \showarticletitle{Adaptive stochastic approximation by the
  simultaneous perturbation method}.
\newblock {\em IEEE Trans. Auto. Ctrl.\/}  {45} (2000), 1839--1853.
\newblock


\bibitem[\protect\citeauthoryear{Styblinski and Tang}{Styblinski and
  Tang}{1990}]%
        {Styblinski_1990_jour_NeuNet}
{M.~A. Styblinski} {and} {T.~S. Tang}. 1990.
\newblock \showarticletitle{Experiments in nonconvex optimization: {S}tochastic
  approximation with function smoothing and simulated annealing}.
\newblock {\em Neural Networks\/} {3}, 4 (1990), 467--483.
\newblock


\bibitem[\protect\citeauthoryear{Suri}{Suri}{1987}]%
        {Suri_1987_jour_JACM}
{R. Suri}. 1987.
\newblock \showarticletitle{Infinitesimal perturbation analysis for general
  discrete event systems}.
\newblock {\it J. ACM} {34}, 3 (1987).
\newblock


\bibitem[\protect\citeauthoryear{Suyari}{Suyari}{2005}]%
        {Suyari_2005_jour_ITTrans}
{H. Suyari}. 2005.
\newblock \showarticletitle{Law of error in {T}sallis statistics}.
\newblock {\em IEEE Trans. Info. Theory\/} {51}, 2 (2005), 753--757.
\newblock


\bibitem[\protect\citeauthoryear{Thistleton, Marsh, Nelson, and
  Tsallis}{Thistleton et~al\mbox{.}}{2007}]%
        {Thistleton_2007_jour_ITTrans}
{W.~J. Thistleton}, {J.~A. Marsh}, {K. Nelson}, {and} {C. Tsallis}. 2007.
\newblock \showarticletitle{Generalized {B}ox-{M}uller method for generating
  {$q$}-{G}aussian random deviates}.
\newblock {\em IEEE Trans. Info. Theory\/} {53}, 12 (2007), 4805--4810.
\newblock


\bibitem[\protect\citeauthoryear{Tsallis}{Tsallis}{1988}]%
        {Tsallis_1988_jour_StatPhy}
{C. Tsallis}. 1988.
\newblock \showarticletitle{Possible generalization of {B}oltzmann-{G}ibbs
  statistics}.
\newblock {\em J. Stat. Phy.\/} {52}, 1-2 (1988), 479--487.
\newblock


\bibitem[\protect\citeauthoryear{Tsallis, Mendes, and Plastino}{Tsallis
  et~al\mbox{.}}{1998}]%
        {Tsallis_1998_jour_PhysicaA}
{C. Tsallis}, {R.~S. Mendes}, {and} {A.~R. Plastino}. 1998.
\newblock \showarticletitle{The role of constraints within generalized
  nonextensive statistics}.
\newblock {\em Physica A\/} {261}, 3--4 (1998), 534--554.
\newblock


\bibitem[\protect\citeauthoryear{Vazquez-Abad and Kushner}{Vazquez-Abad and
  Kushner}{1992}]%
        {Vazquez_1992_jour_AppProb}
{F.~J. Vazquez-Abad} {and} {H.~J. Kushner}. 1992.
\newblock \showarticletitle{Estimation of the derivative of a stationary
  measure with respect to a control parameter}.
\newblock {\em J. App. Prob.\/}  {29} (1992), 343--352.
\newblock


\bibitem[\protect\citeauthoryear{Vignat and Plastino}{Vignat and
  Plastino}{2006}]%
        {Vignat_2006_jour_PhysicaA}
{C. Vignat} {and} {A. Plastino}. 2006.
\newblock \showarticletitle{Poincar{\'{e}}'s observation and the origin of
  {T}sallis generalized canonical distributions}.
\newblock {\em Physica A\/} {365}, 1 (2006), 167--172.
\newblock


\bibitem[\protect\citeauthoryear{Vignat and Plastino}{Vignat and
  Plastino}{2007}]%
        {Vignat_2007_jour_PhyA}
{C. Vignat} {and} {A. Plastino}. 2007.
\newblock \showarticletitle{Central limit theorem and deformed exponentials}.
\newblock {\em J. Phy. A\/} {20}, 45 (2007).
\newblock


\bibitem[\protect\citeauthoryear{Williams}{Williams}{1991}]%
        {Williams_1991_book_Cambridge}
{D. Williams}. 1991.
\newblock {\em Probability with martingales}.
\newblock Cambridge University Press.
\newblock


\end{thebibliography}



\medskip

\end{document}